\documentclass[a4paper,twocolumn,
english,aps,pre,floatfix,groupedaddress,showpacs,nofootinbib]{revtex4-1}
\usepackage[T1]{fontenc}
\usepackage[latin1]{inputenc}
\usepackage{amsmath}
\usepackage{babel}
\usepackage{graphics}
\usepackage{amssymb}
\usepackage{mathrsfs}
\usepackage{dcolumn}


\begin{document}
\title
{Dynamics of driven flow with exclusion in graphene-like structures} 
\author {R. B. \surname{Stinchcombe}}
\email{r.stinchcombe1@physics.ox.ac.uk}
\affiliation{Rudolf Peierls Centre for Theoretical Physics, University of
Oxford, 1 Keble Road, Oxford OX1 3NP, United Kingdom}
\author {S. L. A. \surname{de Queiroz}}
\email{sldq@if.ufrj.br}
\affiliation{Instituto de F\'\i sica, Universidade Federal do
Rio de Janeiro, Caixa Postal 68528, 21941-972
Rio de Janeiro RJ, Brazil}

\date{\today}

\begin{abstract} 
We present a mean-field theory for the dynamics of driven flow
with exclusion in graphene-like structures, 
and numerically check its predictions. We treat first
a specific combination of bond transmissivity rates, 
where mean field predicts, and numerics to a large extent confirms, 
that the sublattice structure characteristic of honeycomb networks becomes
irrelevant. Dynamics, 
in the various regions of the phase diagram set by open boundary 
injection and ejection rates, is then in general identical to that of 
one-dimensional (1D) systems, although some discrepancies remain 
between mean-field theory and numerical results, in similar ways
for both geometries.
However, at the critical point for which the characteristic 
exponent is $z=3/2$ in 1D, the
mean-field value $z=2$ is approached for very large systems with 
constant (finite)  aspect ratio.  
We also treat a second combination of bond (and boundary) rates where,
more typically, 
sublattice distinction persists. For the two rate combinations, in
continuum or late-time limits respectively the coupled sets of mean 
field dynamical equations become tractable with various techniques and 
give a two-band spectrum, gapless in the critical phase. While
for the second rate combination quantitative discrepancies 
between mean field theory 
and simulations increase for most properties and boundary rates investigated, 
theory still is qualitatively correct in general, and
gives a fairly good quantitative account of features 
such as the late-time  evolution of density 
profile differences from their steady state values.

\end{abstract}
\maketitle
 
\section{Introduction} 
\label{intro} 
In this paper we consider the dynamic evolution, as well as selected steady-state
properties, of a generalization of the totally asymmetric simple exclusion 
process (TASEP) to two-dimensional honeycomb structures. A previous 
publication~\cite{hex13} focused mainly on the evaluation of steady-state
currents for several variations of such structures.    

The TASEP, in its one-dimensional (1D) version, exhibits 
many non-trivial properties including flow phase changes, because of its
collective character~\cite{derr98,sch00,mukamel,derr93,rbs01,be07,cmz11}. 
The TASEP and its generalizations have been
applied to a broad range of non-equilibrium physical contexts, from
the macroscopic level such as highway traffic~\cite{sz95} to the microscopic,
including sequence alignment in computational biology~\cite{rb02}
and current shot noise in quantum-dot chains~\cite{kvo10}.
 
In the time  evolution  of the 1D TASEP,
the particle number $n_\ell$ at lattice site $\ell$ can be $0$ or $1$, 
and the forward hopping of particles is only to an empty adjacent site. 
In addition to the stochastic character provided by random selection of site occupation 
update~\cite{rsss98,dqrbs08}, the instantaneous current $J_{\ell\,\ell+1}$ 
across the bond from $\ell$ to $\ell +1$ depends also on the stochastic attempt rate, 
or bond (transmissivity) rate, $p_\ell$, associated with it. Thus,  
\begin{equation}
J_{\ell\,\ell+1}= \begin{cases}{n_\ell (1-n_{\ell+1})\quad {\rm with\ probability}\ p_\ell}\cr
{0\qquad\qquad\qquad {\rm with\ probability}\ 1-p_\ell\ .}
\end{cases}
\label{eq:jinst}
\end{equation}
In Ref.~\onlinecite{kvo10} it was argued that the ingredients of (1D) TASEP
are expected to be physically present in the description of electronic
transport on a quantum-dot chain; namely,
the directional bias would be provided by an external voltage difference
imposed at the ends of the system, and the exclusion effect by on-site Coulomb 
blockade.

Apart from the importance of generalizing fundamental dynamic studies
of the linear chain TASEP to higher-dimensional lattices and structures,
the present work, and in particular its emphasis on honeycomb structures,
is partly motivated
by recent progress in the physics of graphene and its quasi-1D
realizations, such as nanotubes and nanoribbons~\cite{RMP}.
Of course the TASEP, as described above, does not provide a realistic
description of electronic transport in carbon allotropes under an applied bias.
However, in the transport context the lattice topology affects how currents
combine, and how they are microscopically located, whether classical or
quantum. It will be seen that these features show up in the model 
we treat by such effects as the sublattice structure seen e.g. 
in steady states for the uniform hexagonal lattice (Secs.~\ref{sec:pneq2q}
and~\ref{sec:num:pneq2q}), 
and as consequent band-doubling in the corresponding spectrum. 
An interesting effect, 
which we exploit, is the similar behavior arising in topologically trivial
linear systems from alternating bond rates (Sec.~\ref{sec:num:pneq2q}).

Although the model is classical, so it does not display quantum interference
effects,  it is the simplest cooperative driven model, with consequent
qualitative properties reflecting aspects of Coulomb blockade
phenomenology in real experiments.

In Ref.~\onlinecite{hex13} we probed 
for the existence of similar  specific signatures
by examining the behavior
of steady-state currents for nanotubes and nanoribbons, against varying system sizes,
and for diverse combinations of bond transmissity rates, as well as
distinct sets of  boundary conditions along the flow direction, namely periodic
(such as to make the system ring-like) and open (with assorted values for injection 
and ejection rates at the ends, to be recalled in detail below).

The latter case of open systems, with open boundary conditions at the ends, 
is by far the most challenging, richest, and most illuminating one, so it (alone)
is the case here considered. As in Ref.~\onlinecite{hex13}, the
present study makes complementary use of mean field analysis and
numerical simulations.

In Section~\ref{sec:dyn-theo} a mean field theory is presented
which describes the time evolution of ensemble-averaged site
occupations under TASEP rules, and applies both to the two-dimensional 
structures under specific consideration here and to their linear
chain counterparts. Section~\ref{sec:dyn-num} deals with numerical tests
of the theory given in Section~\ref{sec:dyn-theo}.
In Section~\ref{sec:conc}, we summarize and discuss our results.

\section{Mean-field theory}
\label{sec:dyn-theo}

For analytic tractability we shall only consider cases where mean flow direction is 
parallel to one of the lattice directions, and bond rates are independent of coordinate 
transverse to the flow direction. These configurations have no bonds orthogonal to the 
mean flow direction; thus they fall easily within the generalized TASEP description to be 
used, where each bond is to have a definite directionality, compatible with that of 
average flow.

\begin{figure}
{\centering \resizebox*{2.8in}{!}{\includegraphics*{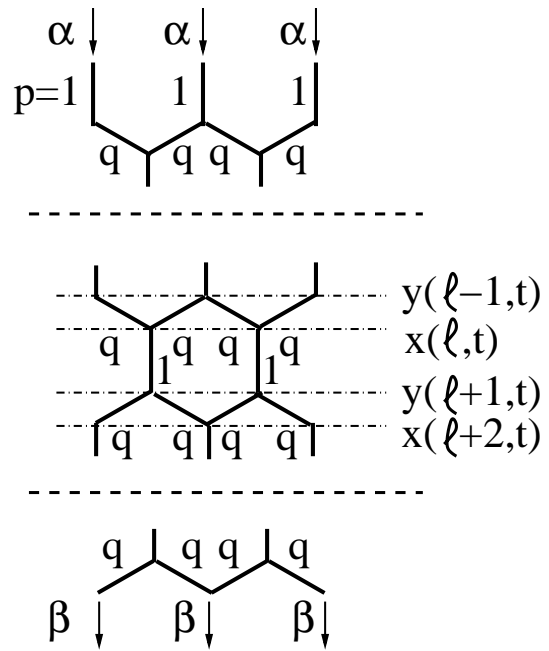}}}
\caption{
Schematic sections of a nanotube, showing (top to bottom):
injection region, midsection, and ejection region. Average flow direction is
from top to bottom of the figure. Bond rates are $p=1$ for bonds
parallel to average flow direction, $q$ otherwise. Injection ($\alpha$)
and ejection ($\beta$) rates are shown next to corresponding (injection and
ejection) sites. 
Periodic boundary conditions across are omitted for clarity.
}
\label{fig:hex}
\end{figure}

Also, we consider structures with an integer number of elementary cells 
(one bond preceding a full hexagon) along the mean flow direction. See
Fig.~\ref{fig:hex}.

From Ref.~\onlinecite{hex13} we have to expect a two-sublattice
character in general, each being of similar character to those for 
chains.
For a special choice of the bond rates defined in Fig.~\ref{fig:hex}
[$\,p=2q$, see the discussion of  Eqs.~(\ref{eq:jmf1})--(\ref{eq:nstrho2})
below$\,$] the steady state sublattices reduce in mean 
field to that of an equivalent uniform-rate chain~\cite{hex13}.  

Throughout this paper only axially symmetric boundary conditions
will be considered, and no rate disorder will be allowed for. Then,
in general, the (mean) dynamic configurations are translationally
invariant in the direction transverse to the tube axis. 
Consistently with this, 
we denote the average occupations
at sites labelled by the longitudinal coordinate $\ell$ ($1 \leq \ell \leq N$) 
by $x(\ell,t)$ and $y(\ell,t)$ with $\ell$ odd and even respectively, corresponding to the
two sublattices (see Fig.~\ref{fig:hex}).

Using mean field factorization, the currents on the two different types
of bond are
\begin{eqnarray}
J_{\ell\,\ell+1}=p\,x_\ell\,(1-y_{\ell+1})\qquad (\ell\ {\rm odd})\ \ \,
\label{eq:jmf1}\\
K_{\ell\,\ell+1}=q\,y_\ell\,(1-x_{\ell+1})\qquad (\ell\ {\rm even})\ .
\label{eq:jmf2}
\end{eqnarray}
Then the general equations for ${\dot x}_\ell$, ${\dot y}_\ell$
at interior sites $\ell$ are
\begin{eqnarray}
{\dot x}_\ell =2 K_{\ell-1\,\ell} - J_{\ell\,\ell+1}\qquad 
(\ell\ {\rm odd})\ \ \,
\label{eq:nstrho1}\\
{\dot y}_\ell = J_{\ell-1\,\ell} - 2 K_{\ell\,\ell+1}\qquad 
(\ell\ {\rm even})\ .
\label{eq:nstrho2}
\end{eqnarray}
From boundary injection and ejection at sites $\ell=1$ and $N$, both on the
$x-$sublattice ($\ell$ odd),  incoming and outgoing currents are
\begin{eqnarray}
\alpha (1-x_1) \equiv J_1
\label{eq:alphabc}\\
2 K_N \equiv x_N\,\beta\ .
\label{eq:betabc}
\end{eqnarray}
In the steady state where ${\dot x}_\ell={\dot y}_\ell=0$, all $\ell$, these
discrete equations specify discrete current balance, making $J_{\ell\,\ell+1}$
and $2K_{\ell\,\ell+1}$ equal and bond-independent 
($={\bar J}$, say), and making
$x_\ell$ and $y_\ell$ reduce to steady state values ${\bar x}_\ell$,
${\bar y}_\ell$, where
\begin{equation}
\alpha\,(1-{\bar x}_1) = {\bar J} = \beta\,{\bar x}_N\ .
\label{eq:alphabetast}
\end{equation}
The distinct steady state sublattice characteristics are seen in 
the (in general)
distinct $\ell-$dependent density profiles  ${\bar x}_\ell$,
${\bar y}_\ell$ which are provided by Mobius map relationships 
between  ${\bar x}_\ell$ and ${\bar y}_{\ell+1}$  resulting from
specified ${\bar J}$ and ${\bar K}$ ($={\bar J}/2$).

From Eqs.~(\ref{eq:jmf1})--(\ref{eq:nstrho2}), it is easy to see
(and was exploited in Ref.~\onlinecite{hex13})
that the sublattice distinction goes away for the special case
$p=2q$. Here the nanotube steady state is that of an equivalent
linear chain, having density profile in general with $\tanh$ or
$\tan$ dependences on $\ell$. 

The value of a continuum approach to the mean field
dynamics of the uniform linear chain is well 
known~\cite{mukamel,rbs01,dqrbs08}, and it exploits a
linearization of the continuum mean field dynamic
equations using the Cole-Hopf transformation~\cite{hopf50,cole51}.
We show in Sec.~\ref{sec:peq2q} that this technique can also be 
successfully used for the nanotube with rates $2q=p=1$ (for 
convenience) and axial symmetry.

In Ref.~\onlinecite{hex13}
it was shown that for the general case $p \neq 2q$, Mobius
maps still apply, from which steady state density profiles are again 
predicted to be
of $\tanh$ or $\tan$ form, but in general different on the two
sublattices. 
Even though on each sublattice separately continuum viewpoints can still
apply (e.g. not too far from critical conditions), standard Cole-Hopf
transformations no longer linearize the coupled nonlinear dynamic equations.
Nevertheless, in Sec.~\ref{sec:pneq2q}, (i) we are there able to use 
another linearization procedure, on the discrete equations for
the dynamics, which gives an asymptotically exact representation
of the mean field dynamics at very late times; and (ii)
furthermore, we can exploit  arguments (see Appendix~\ref{sec:app1}) 
based on the existence of two separate relaxation time scales, from which 
it follows that a continuum-like picture is in fact feasible for not 
very short times.
It will be seen that this, combined with simulation, can give a
particularly clear and direct probe of critical dynamics.

\subsection{Continuum approach for $p=2q=1$}
\label{sec:peq2q}

In this case, no longer needing to distinguish sublattices, 
the notation
$\rho(\ell, t)$ can now be used for the density profile.
The continuum version of the bond current is then
\begin{equation}
J \sim \rho (1-\rho) - \frac{1}{2}\frac{\partial \rho}{\partial \ell}
\label{eq:contcurr}
\end{equation}
from which one arrives at the following form of the steady state
profile:
\begin{equation}
{\bar \rho}= \frac{1}{2} + \frac{1}{2} Z\,\tanh \left[Z (\ell-\ell_0)\right]\ ,
\label{eq:contprof}
\end{equation}
from ${\bar J}=$ constant $=(1-Z^2)/4$, with $Z$ real or pure
imaginary depending on whether the steady state current
is less or greater than the critical value $J_c=1/4$.
The resulting continuum dynamic equation
\begin{equation}
\frac{\partial \rho}{\partial t} =-\frac{\partial}{\partial \ell}
\left[\rho(1-\rho) -\frac{1}{2} \frac{\partial \rho}{\partial \ell}\right]
\label{eq:contdyn}
\end{equation}
is easily reduced to a linear (diffusion) equation for the variable $u$,
by the Cole-Hopf transformation~\cite{hopf50,cole51}
\begin{equation}
\rho - \frac{1}{2}=  \frac{1}{2} \frac{\partial}{\partial \ell}\ \ln u\ .
\label{eq:c-h}
\end{equation}
A general solution reducing as $t \to \infty$ to the steady state
profile $\bar \rho$ given in Eq.~(\ref{eq:contprof}) is
\begin{eqnarray}
u={\bar u} + \Sigma\ ,\qquad {\rm where}\\
\label{eq:usig}
{\bar u} = \cosh \left[Z(\ell-\ell_0)\right]\,e^{\frac{1}{2}Z^2t} \\
\label{eq:u}
\Sigma = \sum_\zeta \left( a_\zeta\,e^{\zeta \ell} +
 a_{-\zeta}\,e^{-\zeta \ell}\right)\,e^{\frac{1}{2}\zeta^2 t}
\label{eq:sig}
\end{eqnarray}
where the sum is over $\zeta^\prime$s, in general complex,
satisfying $\Re\, \zeta^2 < \Re\, Z^2$. For the validity of
the continuum approximation, $Z$ and all effective  $\zeta^\prime$s
arising should be small. The boundary conditions, Eqs.~(\ref{eq:alphabc})
and~(\ref{eq:betabc}), which determine them can be rewritten as
(for all $t$)
\begin{eqnarray}
\rho(0,t)=\alpha\  \,\\
\label{eq:rhoalpha}
\rho(N+1,t)=1-\beta\ ,
\label{eq:rhobeta}
\end{eqnarray}
where $\rho(0,t)$ and $\rho(N+1,t)$ are the extrapolations of the
solution, Eqs.~(\ref{eq:c-h})--(\ref{eq:sig}), of the dynamic
equations to the fictitious sites immediately outside of the
system boundaries.
These then have to be satisfied by the steady state part 
${\bar \rho} =(1/2)+ (1/2)\partial \ln {\bar u}/\partial \ell$, as well
as by the time-dependent parts of the extended $\rho$. 
The requirements on $\bar \rho$
give $Z$, $\ell_0$, in particular requiring $Z$ real for $\alpha < 1/2$
or $\beta < 1/2$; or $Z$ pure imaginary for $\alpha >1/2$
and $\beta > 1/2$. From the time-dependent parts the boundary conditions
then require $\partial \ln \Sigma / \partial \ell$ equal to
$\mu_1 \equiv 2\alpha -1$ and $\mu_2 \equiv 1 -2\beta$ at
$\ell=0$ and $\ell=L \equiv N+1$ respectively. That leads to
\begin{equation}
\frac{a_{-\zeta}}{a_{\zeta}}=\frac{\zeta - \mu_1}{\zeta + \mu_1}=
e^{2\zeta L}\left(\frac{\zeta - \mu_2}{\zeta + \mu_2}\right)\ ,
\label{eq:zetamu}
\end{equation}
giving both the allowed complex wave vectors $\zeta$, and the
ratio of associated amplitudes.
Initial conditions then in principle complete the determination of 
all amplitudes, by the analogue of Fourier analysis.

Some special cases will be of interest in what follows, namely 
$\alpha=\beta$  and $\alpha+\beta=1$.

For $\alpha=\beta$, the open boundary condition restrictions make 
$\ell_0=L/2$, and, for $\alpha=\beta <1/2$, $Z$ is real, 
say $Z \equiv K$,
with $K=1-2\alpha + {\cal O}(e^{-(1-2\alpha)\,L})$ -- 
so the dynamics is
relaxation  to the steady state of the low current phase, having a kink
in the middle of the system; while, for  $\alpha=\beta >1/2$,
$Z$ is pure imaginary, say $Z \equiv iQ$, with $Q=(2/L)((\pi/2)-
\pi/[L(2\alpha -1)])$,  and the relaxation is towards the high current
phase steady state. 

For the critical subcase  $\alpha=\beta =1/2$,
one has $Z =0$, $\zeta_n=n\pi\,i/L\equiv iq_n$, $a_{-\zeta}=
a_\zeta \equiv a_n$. So for this case ${\bar u}=0$ and
$u=\Sigma=\sum_n a_n\,(e^{\zeta_n}+e^{-\zeta_n})\,e^{\frac{1}{2}\zeta_n^2\,t}$
making
\begin{equation}
\rho(\ell,t)=\frac{1}{2} -\frac{1}{2}\frac{\sum_{n=1}^{L} q_n a_n 
\sin (q_n\ell)\,
e^{-\frac{1}{2}q_n^2 t}}{\sum_{n=0}^{L} a_n \cos (q_n\ell)\,
e^{-\frac{1}{2}q_n^2 t}}
\label{eq:rhoab5}
\end{equation}
A given initial profile $\rho(\ell,0)$ would complete the
determination of $\rho(\ell,t)$ by providing the coefficients $a_n$,
by the equivalent of Fourier cosine analysis of $\exp \left[ \int_0^\ell
d\ell^{\,\prime}\,\left(2\rho(\ell^{\,\prime},0) - 1)\right)\right]$ 
in the present case.

For an initially empty lattice, for example, this gives
\begin{equation}
a_n=\frac{2}{L}\frac{\left[ 1-(-1)^n\,\exp(-L/2)\right]}
{1+q_n^2} \approx \frac{2}{L}\,\left[1+q_n^2\right]^{-1}\ .
\label{eq:empty}
\end{equation}
Then, for late times $t \gtrsim (L/\pi)^2$,
\begin{equation}
\rho \approx \frac{1}{2}- \frac{1}{2}\,q_1 \sin \left(\frac{\pi\ell}{L}
\right) \exp\left[-\frac{1}{2}\left(\frac{\pi}{L}\right)^2 t\,\right]
\label{eq:late}
\end{equation} 
while for early times $1 \ll t \ll  (L/\pi)^2$
\begin{equation}
\rho \approx \frac{1}{2}-\frac{\partial}{\partial \ell} \ln I(\ell,t)\ ,
\quad I(\ell,t) =\int_0^\pi d\zeta\,\frac{\cos(\zeta\ell)\,e^{-\frac{1}{2}
\zeta^2t}}{1+\zeta^2}
\label{eq:early}
\end{equation}
making $I(\ell,t)\,\sqrt{t}$ essentially a function of $\ell/\sqrt{t}$,
and $\rho$ linear in $\ell$ ($\rho \approx (1/2)-(\ell/2t)$) up
to $\ell \sim {\cal O}(\sqrt{t})$. This is of course related to the buildup 
of density from the injection site, and is evident in simulation results
shown in Sec.~\ref{sec:dyn-num}, see Fig.~\ref{fig:in_slope}.

For $\alpha+\beta=1$, the boundary restrictions on the steady state
are consistent with $Z=1-2\alpha \equiv \lambda$, and 
$\ell_0 \to \infty$
for $\lambda >0$ and  $\ell_0 \to -\infty$ for $\lambda <0$ (kinks far
outside of the system). From the other boundary restrictions,
$\zeta_n =n\pi\,i/L$, and $a_{-\zeta_n}/a_{\zeta_n}=(\zeta_n+\lambda)/
(\zeta_n-\lambda)$. These make the steady state $\bar u$ proportional
to $\exp\left[-\lambda \ell+\frac{1}{2}\lambda^2t\right]$ and
\begin{equation}
\frac{\Sigma}{\bar u} \equiv S =\sum_{n=1}^{L}\left(a_{\zeta_n}\,
e^{\zeta_n \ell}+a_{-\zeta_n}\,e^{-\zeta_n \ell}\right)\,
e^{\lambda\ell}\,e^{\frac{1}{2}(\zeta_n^2-\lambda^2)t}\ .
\label{eq:S}
\end{equation}
Then the time dependent density profile becomes
\begin{equation}
\rho= \frac{1}{2} +\frac{1}{2}\frac{\partial}{\partial \ell} \ln u=  
\frac{1}{2} +\frac{1}{2}\frac{\partial}{\partial \ell} 
\ln [\,{\bar u}(1+S)\,]=
\alpha+\frac{1}{2}\frac{\partial S/\partial \ell}{1+S}\ .
\label{eq:rhoab1}
\end{equation}
In this case the relaxation is towards the constant (factorizable)
steady state profile ${\bar \rho}_\ell=\alpha$; at very late times
one has
\begin{eqnarray}
\rho-\alpha \approx \frac{1}{2}\frac{\partial S}{\partial \ell}=
\sum_{n=1}^{L}\left(i\lambda -\frac{n\pi}{L}\right)
\,a_{\zeta_n}
\sin \left(\frac{n\pi \ell}{L}\right)\,e^{\lambda \ell}\,\times 
\nonumber\\
\times\,\exp \left[-\frac{1}{2}\left(\lambda^2+(n\pi/L)^2\right)t
\right]\qquad
\ 
\label{eq:rhoab1b}
\end{eqnarray}
where $\lambda=1-2\alpha$. For $\alpha=1/2$, Eqs.~(\ref{eq:rhoab1})
and~(\ref{eq:rhoab1b}) reduce to Eq.~(\ref{eq:rhoab5}).

The late-time results in Eqs.~(\ref{eq:late}) and~(\ref{eq:rhoab1b})
above, and others to be given in Sec.~\ref{sec:pneq2q}
[$\,$especially Eqs.~(\ref{eq:xell2}) and ~(\ref{eq:yell2})$\,$] 
can give guidance beyond the mean field regime used
to obtain them. The correspondence, within mean field, between chain
and nanotube for the case $p=2q$ (given for the steady state in 
Ref.~\onlinecite{hex13} and extended here to dynamics) implies 
the same mean field exponents, and this is seen 
also for $2q \neq p$ below, see Sec.~\ref{sec:pneq2q}. In particular 
the functional dependences on $\ell/\sqrt{t}$ and $t/{L}^2$ seen
above [$\,$in Eq.~(\ref{eq:late}), and in the equivalent
Eq.~(\ref{eq:rhoab1b}) for $\lambda=0\,$]
correspond to the mean field value $2$ of the dynamic
critical exponent $z$. But one can reasonably 
expect the (wide) nanotube
to have different critical exponents from those known for the chain,
since the two have different dimensions.

The simulation method in Sec.~\ref{sec:dyn-num} is able to exhibit
these differences, and the mean field analytic results
suggest a direct method to find them, by
exploiting the late time behavior, in particular by using the
slowest-relaxing mode.

The results in Eq.~(\ref{eq:late}) and~(\ref{eq:rhoab1b}) (the latter,
from just the $n=1$ term) provide mean field examples of that
mode, and suggest that its isolation, by working at late times,
particularly when the system is relaxing to a uniform steady state
[$\,$using $\rho(\ell,t)-{\bar \rho}_\ell\,$], can give the most
unencumbered way of numerically investigating the critical
dynamics. Finite-size scaling using fitting forms for $\rho(\ell,t)-
{\bar \rho}(\ell)$, like in Eq.~(\ref{eq:late}) or in the $n=1$
mode of Eq.~(\ref{eq:rhoab1b}), but with the time-dependent
factor replaced by $\exp [-ct\,{L}^{-z}]$ are suggested: the
general form $f(\ell/L,t/{L}^z)$ could, from the last 
surviving eigenmode of the evolution operator $e^{-Ht}$, 
go over to a factorizable form having
an $e^{-t/\tau}$ time-dependent factor, with $\tau \sim {L}^z$,
and a spatially-dependent factor with nodes near $\ell=0$, $L$
(from boundary conditions) and a symmetric form [$\,$like in 
Eq.~(\ref{eq:late})$\,$] or with an extra factor $e^{\lambda\ell}$
as in  Eq.~(\ref{eq:rhoab1b}), the latter in cases with ${\bar \rho}
\neq 1/2$.
These ideas are exploited in Sec.~\ref{sec:dyn-num}, both for the
chain and for the nanotube.

\subsection{Discrete late-time method, for $p \neq 2q$}
\label{sec:pneq2q}

Here we develop an analytic method for the late time dynamics, 
which is applicable for general rates $\alpha$, $\beta$, $p$,
$q$ where sublattices are distinct and remain so even in the
eventual steady state. Unlike Sec.~\ref{sec:peq2q}
using the continuum approach, the method proceeds from the
discrete mean field dynamic equations and linearizes them
by working to first order in differences of site densities
from steady state values.

The discrete steady state densities are determined by the Mobius
maps introduced in Ref.~\onlinecite{hex13}, which result from
steady state internal current balance, together with boundary 
conditions, as explained after Eq.~(\ref{eq:betabc}). If these
densities are site-dependent the difference dynamical equations
resulting from the linearization procedure have site-dependent
coefficients, making them in general intractable.
For
\begin{equation}
\alpha=2q(1-\beta)\qquad\quad (p \equiv 1)\ 
\label{eq:newcond}
\end{equation}
the steady-state densities 
given by the Mobius mappings can be uniform on each sublattice,
while in general remaining distinct.

The analysis now to be given treats that case, at general $q$,
for which the coupled linear difference equations have constant
coefficients. Their solutions are linear combinations of 
factorizable solutions, involving a secular relation
between the frequency and complex wave vectors involved.
The boundary conditions determine the allowed values
of the complex wave vectors and relationships between
amplitudes of degenerate components.

The uniform steady state density profile values $\bar x$,
$\bar y$ on the two sublattices correspond to fixed points
of the discrete Mobius maps. Such fixed points are directly
available from the basic internal and boundary current balance
equations
\begin{equation}
\alpha (1-{\bar x})={\bar x}(1-{\bar y})=2q{\bar y}(1-{\bar x})=
\beta {\bar x}\ .
\label{eq:curbal2}
\end{equation}
Provided $\alpha=2q(1-\beta)$ these result in
\begin{equation}
{\bar x}=\frac{\alpha}{\alpha+\beta}\ ;\qquad {\bar y}=1-\beta\ .
\label{eq:xbarybar}
\end{equation}
An important subcase to be distinguished and developed later in this
section is the critical situation, where the two fixed points for
each sublattice Mobius map coincide (corresponding to $Z=0$
in the continuum steady state description in Eq.~(\ref{eq:contprof}),
see Sec.~\ref{sec:peq2q}).

Starting from the discrete mean field dynamical Eqs.~(\ref{eq:nstrho1})
and~(\ref{eq:nstrho2}) the linearization procedure, valid for
sufficiently late times, is implemented by inserting $x_\ell={\bar x}+
\delta_\ell$, $y_\ell={\bar y}+\varepsilon_\ell$
and including only terms up to first order in $\delta_\ell$,
$\varepsilon_\ell$.

The zeroth order terms involving only $\bar x$ and $\bar y$ 
are those appearing in the steady state current balance, so
they cancel. The resulting coupled linear difference equations
for the time-dependent $\delta_\ell$, $\varepsilon_\ell$ are solved 
by superpositions of factorizable solutions of the form
\begin{eqnarray}
\delta_\ell =g_\zeta\,\exp(\zeta \ell -\lambda t)\ 
\label{eq:fac1}\\
\varepsilon_\ell =h_\zeta\,\exp(\zeta \ell -\lambda t)\  
\label{eq:fac2}
\end{eqnarray}
for specific $\zeta$-- and $\lambda$--dependent ratios $h_\zeta/g_\zeta$
provided $\zeta$ and $\lambda$ satisfy the secular relation
\begin{equation}
\lambda^2-r\lambda+S(\zeta)=0\ ,
\label{eq:secrel}
\end{equation}
where
\begin{eqnarray}
r=1+2q+(1-2q)({\bar x}-{\bar y})\ ;\nonumber\\
S(\zeta)=S_0-\left(S_+\,e^{\,\zeta}+S_-\,e^{-\zeta}\right)
\label{eq:rsdef}
\end{eqnarray}
with
\begin{eqnarray}
S_0=2q(1-{\bar x}-{\bar y})+4q{\bar x}{\bar y}\nonumber\\
S_+=2q{\bar x}{\bar y}\qquad\qquad\qquad\ \ \, \nonumber\\
S_-=2q(1-{\bar x})(1-{\bar y})\ .\quad\, 
\label{eq:s0pm}
\end{eqnarray}
To fit the boundary conditions at all times it is necessary to
combine degenerate modes, i.e., modes with $\zeta_1 \neq \zeta_2$
such that $\lambda(\zeta_1)=\lambda(\zeta_2)$. A sufficient
condition for this is  $S(\zeta_1)=S(\zeta_2)$, from which
\begin{equation}
e^{\,\zeta_1+\zeta_2}= \frac{S_-}{S_+} \equiv e^{2\phi}\ .
\label{eq:phidef}
\end{equation} 
Then, with $\eta_i \equiv \zeta_i-\phi$, the degeneracy condition
becomes $\eta_1=-\eta_2$. That allows the superposition of degenerate
modes for $\delta_\ell$ to be written as
\begin{equation}
\delta_\ell=\sum_\eta \left(g_{\phi+\eta}\,e^{\eta\ell}+
g_{\phi-\eta}\,e^{-\eta\ell}\right)\,e^{\phi\ell}\,
e^{-\lambda(\eta+\phi)t}\ ,
\label{eq:delta}
\end{equation}
and similarly for $\varepsilon_\ell$ (where $h_{\phi\pm\eta}$
replace  $g_{\phi\pm\eta}$).

The secular relation between $\lambda$ and $\zeta$ can be rewritten
as one between $\lambda$ and $\eta$ using
\begin{eqnarray}
S(\zeta=\eta+\phi)=S_0-{\mathscr S}(\eta)\quad {\rm where}\quad
\nonumber\\
{\mathscr S}(\eta)=\sqrt{S_+\,S_-}\,\left(e^\eta+e^{-\eta}\right)\ .
\qquad\quad
\label{eq:scal}
\end{eqnarray}
For the boundary conditions to be maintained by the full
time-dependent profiles $x_\ell={\bar x}+\delta_\ell$,
$y_\ell={\bar y}+\varepsilon_\ell$, the differences
$\delta_\ell$, $\varepsilon_\ell$ have both to vanish at 
$\ell=0$ and $\ell=L$ at all times. That requires
$g_{\phi+\eta}+g_{\phi-\eta}=0=h_{\phi+\eta}+h_{\phi-\eta}$
and $e^{2\eta L}=1$, so the allowed $\eta$'s are
$\eta_n=\pi n i/L \equiv iq_n$.

Consequently the space- and time-dependent sublattice density
profiles are, to linear order,
\begin{eqnarray}
x_\ell(t)={\bar x}+\sum_n G_n \sin q_n\ell\,e^{\phi\ell}\,
e^{-\lambda_nt}
\label{eq:xt}\\
y_\ell(t)={\bar y}+\sum_n H_n \sin q_n\ell\,e^{\phi\ell}\,
e^{-\lambda_nt}
\label{eq:yt}
\end{eqnarray}
where
\begin{equation}
q_n =\frac{n\pi}{L}\ ,\qquad e^{2\phi}= \left(\frac{1-{\bar x}}{\bar x}
\right)\left(\frac{1-{\bar y}}{\bar y}\right)
\label{eq:qphi}
\end{equation}
and $\lambda_n$ satisfies
\begin{equation}
\lambda_n^2-r\lambda_n+S_0-{\mathscr S}(iq_n)=0
\label{eq:lambdars}
\end{equation}
where $r$ and $S_0$ are given in Eqs.~(\ref{eq:rsdef}) 
and~(\ref{eq:s0pm}), and 
\begin{equation}
{\mathscr S}(iq_n)=4q\,\sqrt{{\bar x}{\bar y}(1-{\bar x})
(1-{\bar y})}\cos q_n\ ,
\label{eq:Siqn}
\end{equation}
with $\bar x$, $\bar y$ given by Eq.~(\ref{eq:xbarybar}).

The coefficients $G_n$ and $H_n$ ($2i\,g_{\phi-\eta}$ and
$2i\,h_{\phi+\eta}$, respectively) are in principle determined
by initial states. For initial states $x_\ell(0)$, $y_\ell(0)$
in the linearization regime, they are the coefficients in the
Fourier sine series for $x_\ell(0)- {\bar x}$,  $y_\ell(0)- {\bar y}$
respectively.

The very late time behavior, from the decay of the last surviving 
time-dependent mode, is described by
$x_\ell(t)- {\bar x}$,  $y_\ell(t)- {\bar y}$ both proportional
to $\sin (\pi\ell/L)\,e^{\phi\ell}\,e^{-\lambda_1t}$,
with $\phi$ from Eq.~(\ref{eq:qphi}) and  
$\lambda_1=\frac{1}{2}\,[r-\sqrt{r^2-4(S_0-{\mathscr S}(iq_1))}]$
from Eqs.~(\ref{eq:qphi})--(\ref{eq:Siqn}).

In general, the distinct sublattices give rise to a two-branch spectrum, which
makes the late-time dynamics for the cases with $2q \neq p$
very different from that with $2q=p$ discussed in 
Sec.~\ref{sec:peq2q}. The spectrum is in general gapped even
in the infinite-system limit ($\lim_{L \to \infty} \lambda_1 >0$)
as a consequence of non-zero $\phi$; the gap goes away 
(as $\phi \to 0$) only in the critical cases, which we now discuss.

The critical steady state has constant (coincident fixed point)
values $x^\ast$, $y^\ast$ for $\bar x$, $\bar y$, related to a
critical current $J_c$ on the bonds with rate $p$, and to critical
boundary rates $(\alpha_c,\beta_c)$ by current balance equations
of type Eq.~(\ref{eq:curbal2}), where each current is $J_c$ such that 
the corresponding sublattice Mobius maps each have coincident fixed 
points. With $p=1$, that requires $[J_c(1-2q)-2q]^2=16q^2\,J_c$,
which makes $x^\ast + y^\ast=1$, hence
\begin{equation}
\phi=0\ ,\qquad S_0=2S_+=2S_-=2q x^\ast y^\ast\ .
\label{eq:xastyast}
\end{equation}
That in turn makes
\begin{equation}
S(\zeta)=S(\eta)=4q x^\ast y^\ast (1-\cosh \zeta)
\label{eq:szeta}
\end{equation}
and the development in Eqs.~(\ref{eq:secrel})--(\ref{eq:Siqn})
simplifies. The results for the time-dependent density profiles
become, to linear order,
\begin{eqnarray}
x_\ell(t)=x^\ast+\sum_n G_n \sin q_n\ell\,e^{-\lambda_nt}
\label{eq:xell}\\
y_\ell(t)=y^\ast+\sum_n H_n \sin q_n\ell\,e^{-\lambda_nt}
\label{eq:yell}
\end{eqnarray}
where
\begin{equation}
\lambda_n=\frac{1}{2}\,\left[ r \pm \sqrt{r^2-16qx^\ast y^\ast
(1-\cos q_n)}\right] \equiv \lambda_\pm (q_n)\ .
\label{eq:lambdan}
\end{equation}
So, $\phi=0$ has produced a gapless spectrum in infinite system
limit, for the critical system, and we now have the analogue
of acoustic and optic modes.

For the finite critical system, the very late behavior of the 
profiles on each sublattice is (using the slowest relaxing "acoustic"
mode, $n=1$, with $\lambda_-$)
\begin{eqnarray}
x_\ell(t)=x^\ast+ G \sin \frac{\pi\ell}{L}\,e^{-\lambda_-(\pi/L)\,t}
\label{eq:xell2}\\
y_\ell(t)=y^\ast+ H \sin \frac{\pi\ell}{L}\,e^{-\lambda_-(\pi/L)\,t}
\label{eq:yell2}
\end{eqnarray}
with
\begin{equation}
\lambda_-\left(\frac{\pi}{L}\right) \sim \frac{S(\zeta)}{r} \sim
\frac{4q x^\ast y^\ast}{r}\,\left(1-\cos \frac{\pi}{L}\right) \propto
\left(\frac{\pi}{L}\right)^2\ .
\label{eq:lambdamin}
\end{equation}
The condition $\alpha=2q(1-\beta)$ for uniform steady state densities,
presently applying, reduces for $q=1/2$ to $\alpha+\beta=1$,
which is a case discussed for general $\lambda=1-2\alpha$
in Sec.~\ref{sec:peq2q}. That case becomes critical for $\lambda=0$,
and then the present formulation with $\phi=0$ together with the
resulting Eqs.~(\ref{eq:xastyast})--(\ref{eq:lambdamin}) all
apply to it (reproducing results in that Section).

A particular important special case is that for the uniform-rate 
nanotube, where $p=q=1$ and $x^\ast=2-\sqrt{2}$, $y^\ast=\sqrt{2}-1$,
$\alpha_c=2(\sqrt{2}-1)$, $\beta_c=2-\sqrt{2}$ (from 
Ref.~\onlinecite{hex13}), agreeing with Eqs.~(\ref{eq:curbal2})
and~(\ref{eq:xbarybar}).

The distinction, one or two bands (from $2q$ equal to $p$ or not)
is a special feature of the nanotube coming from its possible
sublattice character, and shared with the TASEP chain with
alternating bond rates $p$, $2q$, which has equivalent mean field
steady state and dynamics.

\section{Numerics}
\label{sec:dyn-num} 

With open boundary conditions at the ends, a nanotube with $N_r$ elementary cells 
parallel to the flow direction, and $N_w$ transversally, has $N_s^{(n)}=N_w \times (4N_r+1)$
sites and $N_b^{(n)}=N_w \times (6N_r+2)$ bonds (including the injection and ejection ones).   

When dealing with strictly 1D geometries, for ease of pertinent comparisons 
with nanotubes we generally took systems with a number of sites $N=4M+1$,
$M$ being an integer.

Here we shall only use so-called {\em bond update} procedures, defined in Ref.~\onlinecite{hex13} 
and briefly recalled below. For a description of the closely-related {\em site} update process, and 
pertinent comparisons with bond update, see Ref.~\onlinecite{hex13}.
 
For a structure with $N_b$ bonds, an elementary time step consists of $N_b$ sequential bond update attempts, 
each of these according to the following rules: (1) select a bond at random, say, bond $ij$, connecting sites
$i$ and $j$; 
(2) if the chosen bond has an occupied site to its left and an  empty site to its right, then 
(3) move the particle across it with probability (bond rate)  $p_{ij}$.
If the injection or ejection bond is chosen, step (2) is suitably modified to 
account for the particle reservoir (the corresponding bond rate being, respectively,  $\alpha$ or $\beta$). 

Thus, in the course of one time step, some bonds may be selected 
more than once for  examination and some may not be examined at all.
This constitutes the {\it random-sequential update} procedure described
in Ref.~\onlinecite{rsss98},
which is the realization
of the usual master equation in continuous time~\cite{rsss98}.
In our simulations, the goal for 1D
uniform systems is to have numerically-generated profiles approach the 
exact steady-state ones given by the operator algebra
described in Ref.~\onlinecite{derr93}, which are an important baseline in
our work and, as recalled in  Ref.~\onlinecite{rsss98}, correspond to
random-sequential update. For consistency, and ease of comparison
between different sets of results within the paper, we also
use random-sequential update for all other cases, namely honeycomb
geometries and non-uniform 1D systems.
Note that other types of update are possible (e.g., ordered-sequential or
parallel), the resulting steady-state phase diagrams in 1D 
being similar in all cases
(but not identical: even the average stationary current differs in either
case, see  Table 1 in Ref.~\onlinecite{rsss98}). 

For specified initial conditions, we generally took ensemble 
averages of local densities and/or currents over $10^6$--$10^7$
independent realizations of stochastic update up to a suitable time 
$t_{\rm max}$, for each of those collecting system-wide samples at 
selected times.

For uniform 1D systems and nanotubes with $p=2q$, the exact steady-state density profiles $\{ {\bar x}_\ell \}$, known 
in 1D for any $\alpha$, $\beta$, and $N$~\cite{derr93} are used as a baseline from which to subtract our late-time 
simulational results $\{ x_\ell(t) \}$, thus focusing  on the evolution of difference profiles 
$\delta x_\ell(t) \equiv x_\ell(t)- {\bar x}_\ell $. 
For nanotubes with $p \neq2q$, or chains with non-uniform rates, both cases 
considered in Sec.~\ref{sec:num:pneq2q}, no such guidance is available. One must then resort to numerically-generated 
steady state profiles. 
 
\subsection{$p=2q$, $\alpha=\beta=1/2$}
\label{sec:num_peq2qab5}

\begin{figure}
{\centering \resizebox*{3.3in}{!}{\includegraphics*{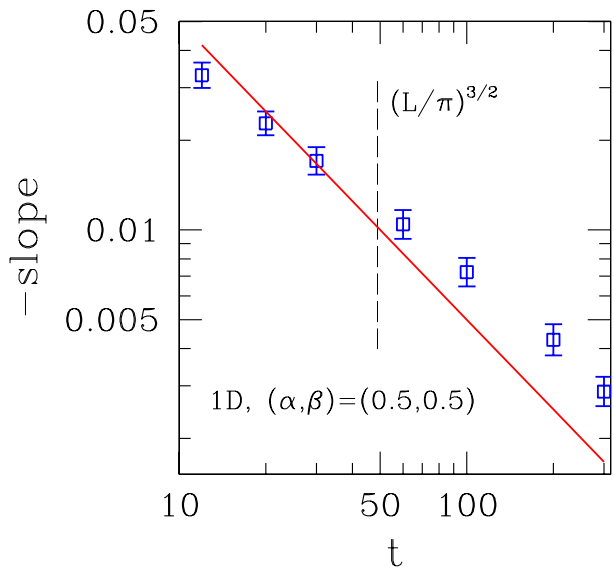}}}
\caption{
Linear chain with $N=41$ sites ($L=N+1$), $\alpha=\beta=1/2$. Double-logarithmic
plot of (negative) initial slopes ($S$) of short-time density profiles 
against time $t$  (points). Continuous line is the mean-field prediction
$S=1/(2t)$, see Eq.~(\protect{\ref{eq:earlyab5}}). The vertical dashed
line indicates the approximate limit of validity of the short-time regime
(see text).
}
\label{fig:in_slope}
\end{figure}

We started by checking the predictions given in Sec.~\ref{sec:peq2q} for  
the  time-dependent density profiles of a 1D system starting from an 
empty lattice.  Eq.~(\ref{eq:early}) predicts 
that for short times $t \ll (L/\pi)^2$,
\begin{equation}
\rho (\ell,t) \approx \frac{1}{2}-\frac{\ell}{2t}
\label{eq:earlyab5}
\end{equation}
near the injection edge, up to $\ell \sim {\cal O}(\sqrt{t})$. For a chain 
with $N=41$ sites, we evaluated the initial slope $\partial \rho/\partial
\ell \vert|_{\ell=0}$ at assorted short times, from straight-line fits
of ensemble-averaged densities at the three leftmost sites. Results are
shown in Fig.~\ref{fig:in_slope}. One sees that agreement between theory and
numerics is rather satisfactory, especially if, drawing on the last 
two paragraphs of Sec.~\ref{sec:peq2q}, and on previous knowledge of
the anomalous scaling for 1D systems at $(\alpha,\beta)=(1/2,1/2)$,
one restricts oneself to data for $t \lesssim (L/\pi)^{3/2}$ 
[$\,$as opposed to $t \lesssim (L/\pi)^2$ from the mean-field picture
leading to Eq.~(\ref{eq:early})$\,$].

\begin{figure}
{\centering \resizebox*{3.3in}{!}{\includegraphics*{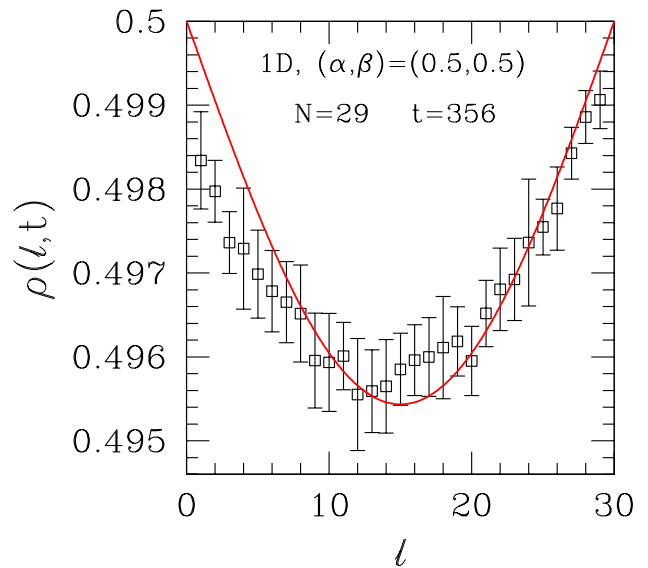}}}
\caption{
Linear chain with $N=29$ sites, $\alpha=\beta=1/2$. Plot of late-time
density profile, starting with an empty lattice at $t=0$. 
Continuous line is the fit to a sine form, 
see Eq.~(\protect{\ref{eq:late}}) and text.
}
\label{fig:sinefit}
\end{figure}

Next we checked the late-time behavior, both for 1D systems and for
nanotubes. Fig.~\ref{fig:sinefit} shows a fit of 
Eq.~(\ref{eq:late}) to the ensemble-averaged density profile for a
1D system, starting from an empty lattice at $t=0$. While the quality
of fit is good, with $\chi^2$ per degree of freedom ($\chi^2_{\rm dof}$) 
equal to $1.35$, one sees that small systematic deviations still remain
near the left (injection) edge. Going over to later times in order to
evince the suppression of such deviations would necessitate
much narrower error bars (since one would be analyzing profiles
much closer to the asymptotic regime), and consequently much longer 
simulations, than in our current setup.

Nevertheless, we now show that it is possible to extract rather 
accurate estimates of
the dynamic exponent $z$ from our data in present form, by once again
referring to the ideas sketched in the last two paragraphs of
Sec.~\ref{sec:peq2q}. Specifically we rewrite Eq.~(\ref{eq:late})
as
\begin{equation}
\rho (\ell,t)=\frac{1}{2}-a^{\,\prime}(L) \sin \left(\frac{\pi\ell}{L}\right)
\exp\left\{-c\,\frac{t}{L^z}\right\} \ ,
\label{eq:late2}
\end{equation}
i.e., while assuming factorization of the $\ell$ and $t$ dependences,
we allow $z$ to be a variable parameter.
For fixed $L$ and a set of suitable $t$ values,
fitting numerically-generated profiles to the
sine dependence in Eq.~(\ref{eq:late2}) produces a 
sequence  of estimates of
\begin{equation}
a^\ast (L,t) \equiv a^{\,\prime}(L)\,\exp\left\{-c\,\frac{t}{L^z}\right\} \ ;
\label{eq:astar}
\end{equation}
the latter set is then fitted to 
\begin{equation}
a^\ast (L,t)=a_0(L) \exp\{-c{\,^\prime} (L)\,t\}\ ,
\label{eq:cprime}
\end{equation}
with $a_0(L)$, $c{\,^\prime} (L)$ as fitting parameters.
Finally, varying $L$ one fits the  corresponding sequence of $c^{\,\prime}(L)$
to a power-law in $L$, thus extracting $z$.

We proceeded as just outlined for: (i) 1D systems, starting with an empty
lattice; (ii) 1D systems, starting with a "sine-like" profile, i.e.,
\begin{equation}
n_\ell(0)=\begin{cases}{1 \quad \ell \leq \frac{N}{4}\ {\rm or}\
 \ell \geq \frac{3N}{4}}\cr
{0 \quad  \frac{N}{4} < \ell < \frac{3N}{4}}
\end{cases}\ ,
\label{eq:sinelike}
\end{equation}
in order to check how sensitive the small late-time systematic deviations, 
referred to above, were to the choice of initial condition;
(iii) nanotubes with $N_w=14$ elementary cells across and varying 
length $N_r$; and finally (iv) nanotubes  with $N_w=N_r$ cells, i.e.
aspect ratio equal to unity. In the latter two cases, sine-like
initial profiles were used.

For (i)--(iii)
we took $N=29$, $41$, $53$, and $69$ (corresponding, for nanotubes, to 
$N_r=7$, $10$, $13$, and $17$) and, for each of these, five $N$- (or $L$)-dependent 
values of $t$ in the late-time approach to steady state. We found that
using a sine-like profile as initial condition does slightly improve the 
quality of profile fits to Eq.(~\ref{eq:late}). For example, in the
corresponding case to that illustrated in Fig.~\ref{fig:sinefit},
we found  $\chi^2_{\rm dof}=0.91$, about a third less than for
an empty-lattice start. 

\begin{figure}
{\centering \resizebox*{3.3in}{!}{\includegraphics*{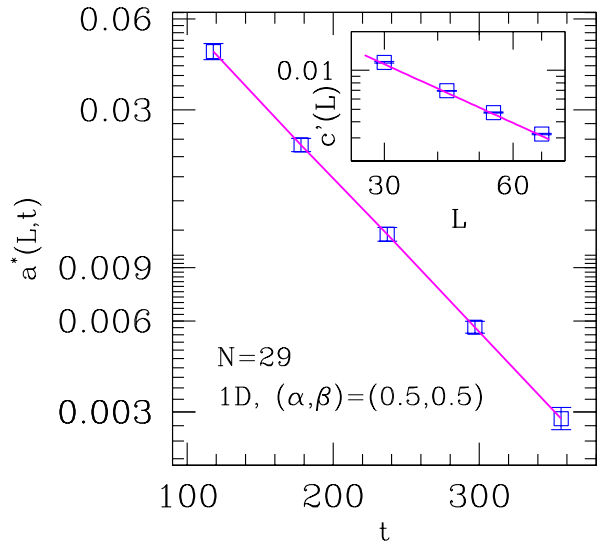}}}
\caption{
Main diagram: log-linear plot of $a^\ast (L,t)$ of
Eq.~(\protect{\ref{eq:astar}}) against $t$ for linear chain with $N=29$ sites. 
The continuous line connects 
numerically-obtained points. Initial condition: sine-like. Inset: double-logarithmic plot of
$c^{\,\prime}(L)$ of Eq.~(\protect{\ref{eq:cprime}}) against $L \equiv N+1$.
The continuous line is a fit of data to $c^{\,\prime}(L) \sim L^{-z}$,
with $z=1.51$. Initial condition: empty lattice.
}
\label{fig:expfit}
\end{figure}

By following the fitting procedures
delineated above our final results were \
$z=1.51(1)$ 
in case (i), 
$z=1.54(1)$ 
in case (ii). The  main diagram in Fig.~\ref{fig:expfit} 
illustrates how well the numerically-evaluated coefficients  
$a^\ast (L,t)$ follow an exponential decay in time. That, as well
as the smooth power-law fit of $c^{\,\prime}(L)$ against $L$ 
shown in the inset, gives strong support
to the ansatz described in Eqs.~(\ref{eq:late2})--~(\ref{eq:cprime}). 

\begin{figure}
{\centering \resizebox*{3.3in}{!}{\includegraphics*{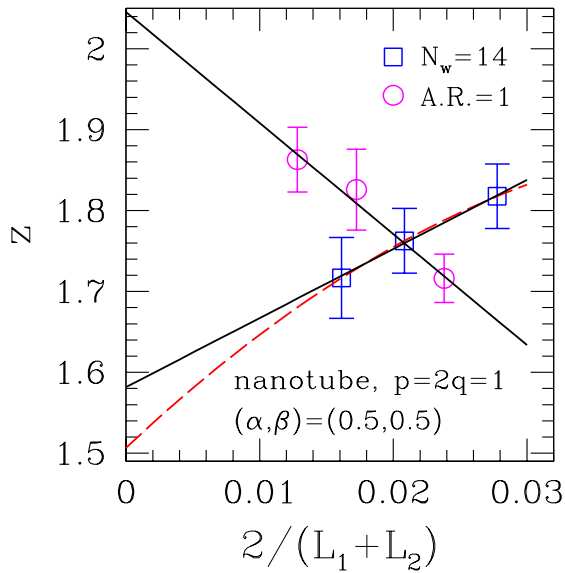}}}
\caption{
Nanotube with $p=2q=1$ at critical point
$(\alpha,\beta)=(1/2,1/2)$.
Points are estimates of dynamical exponent $z$ resulting from
two-point fits of $c^{\,\prime}(L)$ 
in Table~\ref{table:1} for pairs of successive lengths
$(L_1,L_2)$, against
$2/(L_1+L_2)$. Squares: fixed width $N_w=14$. Circles:
fixed aspect ratio (A. R.)$=1$.
Full lines: linear fits. Dashed line: parabolic
fit [$\,$for $N_w=14$ only$\,$] (see text). Initial condition: 
sine-like in all cases.}
\label{fig:ntfit}
\end{figure}

Analysis of case (iii) for the nanotube produced a less clear-cut picture
concerning the final estimate of $z$. Although the exponential decay in time
of the $a^\ast (L,t)$ still holds to excellent accuracy, resulting in the
coefficients $c^{\,\prime}(L)$  listed under the heading $(a)\ N_w=14$
in Table~\ref{table:1}, a 
single power-law fit of the latter against $L$ gives $z=1.76(2)$. By drawing on
ideas for successively iterating sequences of finite-size approximants of
quantities of interest~\cite{bn}, we produced a set of two-point fits of
data for pairs $(L_1,L_2)=(30,42)$, $(42,54)$, and $(54,70)$. Plotting
such set against $2/(L_1+L_2)$,  we arrived at the
following extrapolated values for $2/(L_1+L_2) \to 0$: $z=1.58(1)$
for a linear fit, $z=1.51(2)$ for a parabolic fit, see Fig.~\ref{fig:ntfit}.

In case (iv) we took $N_r=N_w=8$, $12$, $16$, and $22$. The sequence
of coefficients $c^{\,\prime}(L)$, obtained along the same lines already 
described, is given in Table~\ref{table:1}, under $(b)$\ Aspect Ratio$=1$.
As shown in Fig.~\ref{fig:ntfit}, by iterating  two-point fits for pairs of 
successive lengths ones gets an 
increasing sequence of estimates of $z$ against increasing $L$.
A straight-line fit gives an extrapolated $z=2.04(4)$. So this indicates
that, while keeping $N_w>1$ fixed one gets essentially one-dimensional
(critical) behavior, allowing for a constant aspect ratio of order unity
one picks (asymptotically) the true two-dimensional dynamics. 
Furthermore, numerics indicate that the latter is  
characterized by the mean field exponent $z=2$.

\begin{table}
\caption{\label{table:1}
For nanotubes with $p=2q=1$, 
$(\alpha,\beta)=(1/2,1/2)$, late-time coefficients 
$c^{\,\prime}(L)$ of Eq.~(\ref{eq:cprime}), 
obtained by the fitting procedure described in the text, for varying system
lengths $L$. $(a)$: fixed width $N_w=14$ hexagons; $(b)$ aspect ratio $=1$.
}
\vskip 0.2cm
\begin{ruledtabular}
\begin{tabular}{@{}ccc}
$\ \ \ L$ &\ &$c^{\,\prime}(L)$\ \ \\
\hline\noalign{\smallskip}
$\ $&$\ (a)\ N_w=14$ &\ \ \\
\hline\noalign{\smallskip}
$\ \ \ 30$ &\ & \ $\,0.00859(12)$\ \  \\
$\ \ \ 42$ &\ & $0.00458(2)$\ \   \\
$\ \ \ 54$ &\ & $0.00291(3)$\ \   \\
$\ \ \ 70$ &\ & $0.00185(2)$\ \   \\
\hline\noalign{\smallskip}
$\ $&$\ (b)\ \ {\rm Aspect\ Ratio}=1$ &\ \ \\
\hline\noalign{\smallskip}
$\ \ \ 34$ &\ & $0.00670(2)$\ \  \\
$\ \ \ 50$ &\ & $0.00340(3)$\ \   \\
$\ \ \ 66$ &\ & $0.00203(1)$\ \   \\
$\ \ \ 90$ &\ & $0.00113(1)$\ \   \\
\end{tabular}
\end{ruledtabular}
\end{table}
 
Going back to the data for fixed $N_w$, for the nanotube 
with $p=2q=1$, $(\alpha,\beta)=(1/2,1/2)$ 
there appears to be a slow crossover towards $z=3/2$ behavior against
increasing system size, which does not have a parallel in strictly
1D systems. 

\begin{figure}
{\centering \resizebox*{3.3in}{!}{\includegraphics*{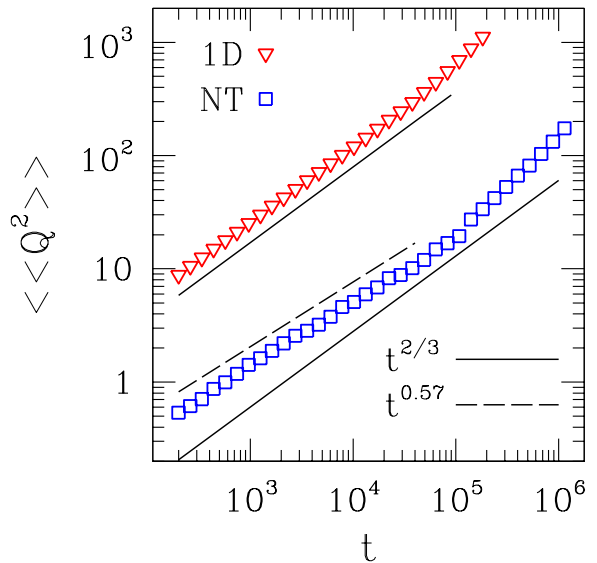}}}
\caption{
Points represent numerically-evaluated second cumulant 
$\langle\langle Q^2 (t)\rangle \rangle$ 
of integrated steady-state current versus time $t$, for 
$(\alpha,\beta)=(1/2,1/2)$. 1D: linear chain, $N=600$ 
(adapted from Ref.~\onlinecite{dq12}). NT:  nanotube of width $N_w=12$ 
hexagons, $N=41$, with bond rates $p=2q=1$. 
Lines indicate power-law dependence
with exponents as shown (see text).
}
\label{fig:ntqcab05}
\end{figure}

We have checked this scenario by investigating a 
steady-state quantity which is well-known to display signatures of
anomalous scaling, namely the cumulants of the integrated 
current~\cite{kvo10,dq12}.  Denoting by $J$ the steady-state average
current through a specified bond, say the one linking sites
$\ell$ and $\ell+1$, and $J_{\ell\,\ell+1}(t^\prime)$ its instantaneous value, 
the associated integrated charge is
${\widetilde Q}_{\ell\,\ell+1}(t) \equiv \int_0^t J_{\ell\,\ell+1} (t^\prime)\,dt^\prime$.
Usually one removes the linear term, and considers 
\begin{equation}
Q(t) \equiv {\widetilde Q}(t)-Jt\ ,
\label{eq:qtilde}
\end{equation}
so $\langle Q(t) \rangle \equiv 0$. For 1D TASEP at 
($\alpha,\beta)=(1/2,1/2)$ the second-order cumulant
$\langle\langle Q^2 \rangle \rangle$  of the integrated current
has been shown~\cite{kvo10,dq12} to exhibit anomalous scaling,
i.e., $\langle\langle Q^2 (t)\rangle \rangle \sim t^{1/z}$ with
$z=3/2$ along a time "window" of width determined by system size
("normal" scaling would correspond to 
$\langle\langle Q^n (t)\rangle \rangle \sim t$ for all $n$). 
In Fig.~\ref{fig:ntqcab05} we show data for both 1D systems,
and for a nanotube with $N_w=12$, $N_r=10$ ($N=41$). 
The apparent behavior $\propto t^{0.57}$ exhibited
for $200 \lesssim t \lesssim 5 \times 10^4$ by the latter
is consistent with $\langle\langle Q^2 (t)\rangle \rangle \sim t^{1/z}$,
using the effective exponent $z=1.76(2)$ found from a global analysis  
of the $c^{\,\prime}(L)$ for fixed $N_w$ of Table~\ref{table:1}.

Still for the nanotube, one can see behavior compatible with $\langle\langle Q^2 
(t)\rangle \rangle \sim t^{2/3}$
for $5 \times 10^4 \lesssim t \lesssim 2 \times 10^5$, until it
crosses over to "normal" scaling $\langle\langle Q^2 (t)\rangle \rangle \sim t$
(of course the latter also takes place for 1D systems, see the corresponding
data in Fig.~\ref{fig:ntqcab05}). The narrowness of the $t^{2/3}$ "window"
is most likely related to the relatively small (longitudinal) system size 
$N$~\cite{kvo10,dq12}.

So in the quasi-one dimensional limit for the $p=2q$ nanotube
at criticality,
the evidence provided both by dynamics (from the scaling of
the $c^{\,\prime}(L)$  of Eq.~(\ref{eq:cprime}) against $L$)
and steady-state (from the scaling of 
$\langle\langle Q^2 (t)\rangle \rangle$ against $t$) consistently
points to an apparent $z\simeq 1.76$ for relatively short systems, and/or
short times (the latter, after full onset of the steady-state regime), 
followed by a crossover towards $z=3/2$ in this case. 

\subsection{$p=2q$, $\alpha+\beta=1$}
\label{sec:num_peq2qa3b7}

For $\alpha+\beta=1$, away from the critical point which was the subject
of Sec.~\ref{sec:num_peq2qab5}, we took a point in the low current
phase of 1D TASEP, namely $\alpha=0.3$, $\beta=0.7$.

\begin{figure}
{\centering \resizebox*{3.3in}{!}{\includegraphics*{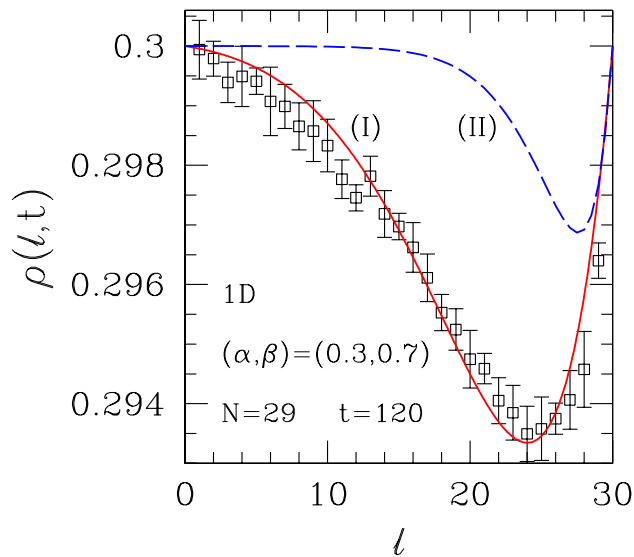}}}
\caption{
Linear chain with $N=29$ sites, $\alpha=0.3$, $\beta=0.7$, $p=2q=1$. 
Plot of late-time
density profile, starting with an empty lattice at $t=0$. 
Curve (I) is a fit to a sine plus exponential form, with $a$
and $b$ of Eq.~(\protect{\ref{eq:a3b7late}}) as adjustable parameters; 
curve (II) is the prediction from Eq.~(\ref{eq:rhoab1b}), adjusted to the empty-lattice 
initial condition, and using only $n=1$; see text.
}
\label{fig:expsinefit}
\end{figure}

Considering 1D systems, starting from an empty lattice, we adapted
Eq.~(\ref{eq:rhoab1b}) for very late times such that only the $n=1$
term in that Equation still survives. In order to investigate
density profiles in this regime we write:
\begin{equation}
\rho(\ell,t)=0.3-a(L,t)\,\sin\left(\frac{\pi\ell}{L}\right)\,e^{b\ell}\ ,
\label{eq:a3b7late}
\end{equation}
where $a(L,t)$ incorporates the exponential time dependence in 
Eq.~(\ref{eq:rhoab1b}), and the factor $b$ in 
Eq.~(\ref{eq:a3b7late}) is predicted to be $b=\lambda=0.4$. In Fig.~\ref{fig:expsinefit}, 
for 1D TASEP  with $N=29$, curve (I) [$\,$full red line$\,$] shows the best fit of 
Eq.~(\ref{eq:a3b7late}) to the simulational results given there, corresponding to
$a=35(4)\times 10^{-5}$, $b=0.144(6)$, with $\chi^2_{\rm dof}=1.8$.
Curve (II) [$\,$dashed blue line] is the prediction of Eq.~(\ref{eq:rhoab1b})
with $\lambda=0.4$, with the $\{a_{\zeta_n}\}$ adjusted to an empty-lattice initial
condition and using only the $n=1$ term.

Although their overall shape is similar, curves (I)  and (II)
significantly differ in (a) the depth and, to a lesser extent, location,
of the minimum on the right-hand side, and (b) the nearly-horizontal segment stretching 
almost midway through the system, exhibited by curve (II), which has no counterpart in
curve (I). While making  $t \approx 110$ in Eq.~(\ref{eq:rhoab1b}), instead of 
"simulation time" $t=120$ reproduces the minimum value shown by numerical data 
(its location, however, remaining unchanged within one lattice spacing), 
point (b) is a permanent feature of the theoretical prediction which reflects the
large value of $\lambda=0.4$ in the profile's exponential $\ell-$dependence in
Eq.~(\ref{eq:rhoab1b}).

The discrepancy between the optimally adjusted
value of the exponential prefactor $b$ of Eq.~(\ref{eq:a3b7late}), on the one hand,
and the theoretical prediction of $\lambda=1-2\alpha$ on the other, is
undoubtedly significant. This indicates that, although simple adaptations enable it to
give an accurate description of the critical systems of Sec.~\ref{sec:num_peq2qab5},
the mean-field theory given above does not quantitatively account for the effects of a
characteristic inverse length $\lambda \neq 0$ in a similarly straightforward way. 
We have found~\cite{unpub} that 
a formulation including the effects of stochastic domain-wall hopping~\cite{ksks98,ps99,ds00}   
on early- and late-time profiles can account for most of the quantitative mismatches between
mean-field theory predictions and simulational results for non-critical cases. 

However, in the present work we limit ourselves to analysing the extent to which the
mean field theory of Sec.~\ref{sec:dyn-theo} can provide useful clues to the actual
behavior of numerically-generated samples. Thus, here we attempt a procedure similar to that 
described in  Sec.~\ref{sec:num_peq2qab5} for extraction of the dynamical exponent. 

In addition to 1D systems, and similarly to Sec.~\ref{sec:num_peq2qab5},
we considered nanotubes both (1) with $N_w=14$ elementary cells across and 
varying  length $N_r$, and  (2) with unit aspect ratio ($N_w=N_r$ cells). 
The time dependences predicted respectively in Eqs.~(\ref{eq:late}), related to the critical ("gapless")
phase  and~(\ref{eq:rhoab1b}) for the "massive" or "gapped" phase,
differ in that the decay rate in the latter has an $L-$independent term, the gap 
[$\,$equal to $\lambda^2/2\,$], related to the characteristic inverse length
$\lambda$.
 
In an attempt to give similar relative 
importance, when compared to the gap contribution, to the finite-size 
dependence to the exponential time decay 
we used $N_r=2$, $3$, $4$, and $5$, corresponding
to $N=9$, $13$, $17$, and $21$ sites.
 
Again, we generated each $a(L,t)$ from adjusting late-time profiles
to Eq.~(\ref{eq:a3b7late}), by allowing both $a$ and $b$ there to vary.
We saw that the fitted value of $b$ generally stayed between  $0.15$ and $0.28$.

We then fitted sequences of varying-$L$ data for $a(L,t)$ to the
$n=1$ term of Eq.~(\ref{eq:rhoab1b}), i.e., 
$a(L,t)=a_0\,\exp(-c_0(L)\,t)$, with $c_0(L)=c+d/L^z$.

\begin{figure}
{\centering \resizebox*{3.3in}{!}{\includegraphics*{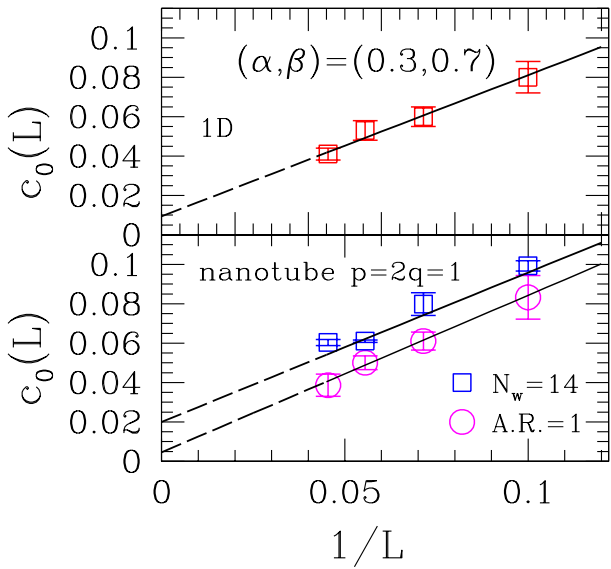}}}
\caption{
For systems with $\alpha=0.3$, $\beta=0.7$, plots of $c_0(L)$ against $1/L$, with $c_0(L)$
defined via $a(L,t)=a_0\,\exp(-c_0(L)\,t)$,
the $a(L,t)$ being given by fitting Eq.~(\ref{eq:a3b7late}) to late-time profiles.  
Upper diagram: 1D systems. Lower diagram: nanotubes with $p=2q=1$; squares: fixed width $N_w=14$; 
circles: fixed aspect ratio (A. R.)=1. }
\label{fig:a3b7ntfit}
\end{figure}

Allowing $z$ to vary freely gave a large amount of scatter 
($0.5 \lesssim z \lesssim 3.5$) among 
fits of four-$L$ data for the three different geometries [$\,$chains, and
nanotubes with either $N_w=14$ or unit aspect ratio$\,$]. 
We then recalled that, for 1D systems in the low-current phase 
$\alpha <1/2$ or $\beta <1/2$ (except on the coexistence line $\alpha=\beta<1/2$)
the effective exponent governing the approach to steady state 
is $z^\prime=1$~\cite{dqrbs08}. This is in contrast to the result from a
rigorous Bethe ansatz calculation~\cite{ess05}, namely $z=0$, and can be explained 
by a mean-field continuum formulation related to kinematic-wave 
propagation~\cite{dqrbs08}. Thus we plotted our data for $c_0(L)$ against $1/L$,
i.e. keeping $z^\prime=1$ fixed. The results are shown in Fig.~\ref{fig:a3b7ntfit}.
It is seen that the numerical data for the sequences of $c_0(L)$ fall reasonably well
onto a straight line consistent with $z=1$, for all three geometries considered.
From the vertical axis intercepts one gets respectively $c=0.009(5)$ (1D), 
$c=0.02(1)$ (nanotube with $N_w=14$), and $c=0.0046(44)$ (nanotube with unit aspect ratio).
These are all definitely much lower than the mean-field prediction $\lambda^2/2=0.08$.
It seems plausible from these data that the gap will vanish for very large nanotubes with finite 
aspect ratio (remaining finite in the quasi- and strictly 1D cases). However, 
the relatively poor quality of the fits [$\,\chi^2_{\rm dof}=0.36$, $10$ and $0.14$, listed for 
each geometry in the same order as the $c$ values$\,$] indicates that a statement of this sort 
would have to be tested more extensively. 

\subsection{$p  \neq 2q$} 
\label{sec:num:pneq2q}

Initially we investigate the nanotube with
$p=q=1$ at $\alpha_c=2(\sqrt{2}-1)$, $\beta_c=2-\sqrt{2}$. These 
rates satisfy the conditions specified in
Eqs.~(\ref{eq:newcond})--~(\ref{eq:xbarybar}), for which                       
the Mobius mapping predicts uniform steady state densities on each 
sublattice, though
in general they remain distinct; namely, in this case they are 
$x^\ast=2-\sqrt{2}$, $y^\ast=\sqrt{2}-1$.

For comparison, we consider also the chain with alternating bond rates
$p$, $2q$ with $p=q=1/2$. The mean-field Mobius mapping for this case 
coincides with that for the $p=q=1$ nanotube, provided the 
injection/ejection rates are suitably renormalised, i.e.
$\alpha=\sqrt{2}-1$, $\beta=1-\sqrt{2}/2$. The respective steady
state sublattice densities are then predicted to coincide, though
of course the rate of approach to steady state on the alternate-bond 
chain is half that for the nanotube.

We took $N_w=14$, $N_r=10$ for the nanotube, and $N=41$ sites for the chain
so both have the same number of sites along the flow direction. For the
remainder of this Section, in both
cases we always started with an empty lattice.

\begin{figure}
{\centering \resizebox*{3.3in}{!}{\includegraphics*{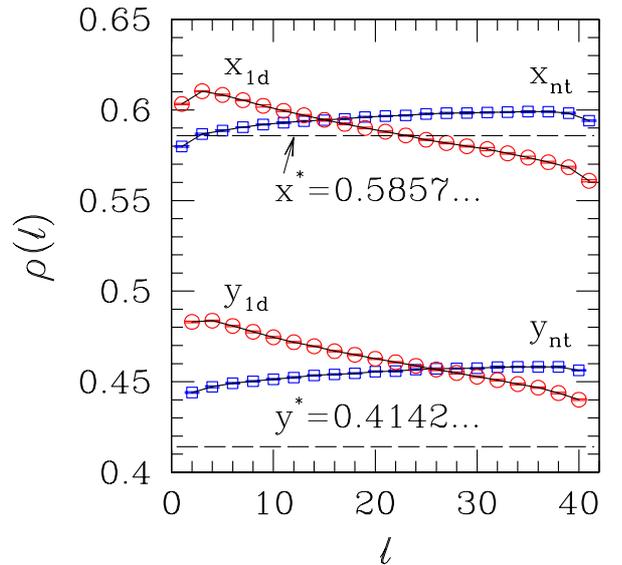}}}
\caption{
Steady state sublattice densities against position along flow direction
for nanotube $p=q=1$ (squares: $x_{nt}$, $y_{nt}$) at $\alpha=2(\sqrt{2}-1)$, 
$\beta=2-\sqrt{2}$,
and for chain with alternating bond rates $p$, $2q$ with $p=q=1/2$ (circles:
$x_{1d}$, $y_{1d}$)
at $\alpha=\sqrt{2}-1$, $\beta=1-\sqrt{2}/2$ (see text). Horizontal dashed
lines show mean field predictions applying for both cases: $x^\ast=2-\sqrt{2}$,
$y^\ast=\sqrt{2}-1$.
}
\label{fig:pq1altc}
\end{figure}

Fig.~\ref{fig:pq1altc} shows that the mean field prediction
of flat sublattice density profiles in steady state is not
fulfilled in numerical simulations. Also, the sublattice profiles for 
the nanotube and the alternating-bond chain do not coincide, at
variance with the fact that they share the same description via
mean-field mapping.
However, the mean field mapping predicts the steady-state sublattice 
densities to within at most $4\%$
(for $x^\ast$) or $16\%$ (for $y^\ast$) of numerical results. Since the 
predicted densities are themselves separated by just over $40\%$, one
can unequivocally ascribe each predicted sublattice profile to the correct 
numerically-generated subset of results.

We defer further discussion of such discrepancies, and others which 
also pertain to steady-state aspects, to Sec.~\ref{sec:fac} below. 
For the moment we investigate, for nanotubes with $p=q=1$, the very
late time behavior of the density profiles. Allowing for the
observed non-uniformity of their limiting steady-state shapes, 
Eqs.~(\ref{eq:xell2}) and~(\ref{eq:yell2}) for a system at criticality 
should translate into:
\begin{eqnarray}
x_\ell(t)=x_\ell^\ast+ G^\prime \sin \frac{\pi\ell}{L}\,e^{-\lambda_-(\pi/L)\,t}
\label{eq:xellnu}\\
y_\ell(t)=y_\ell^\ast+ H^\prime \sin \frac{\pi\ell}{L}\,e^{-\lambda_-(\pi/L)\,t}
\label{eq:yellnu}
\end{eqnarray}
where now the position-dependent $x_\ell^\ast$, $y_\ell^\ast$ are to
be numerically obtained from steady-state simulation data.

\begin{figure}
{\centering \resizebox*{3.3in}{!}{\includegraphics*{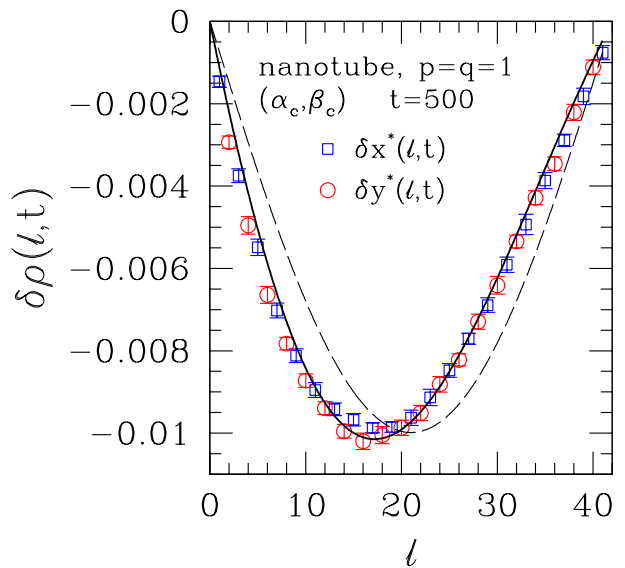}}}
\caption{
Late-time difference profiles, $\delta x_\ell(t) 
\equiv x_\ell(t)-x_\ell^\ast$, and similarly for $\delta y_\ell(t)$,
against position along flow direction
for nanotube $p=q=1$ at $\alpha_c=2(\sqrt{2}-1)$, $\beta_c=2-\sqrt{2}$,
for $t=500$. The dashed line is the fit of  $\delta x_\ell(t)$ to a sine form,
see Eqs.~(\ref{eq:xellnu}),~(\ref{eq:yellnu}). The full line is a fit
of  $\delta x_\ell(t)$ to a sine-plus-exponential form, see Eq.~(\ref{eq:abfit})
and text.
}
\label{fig:pq1diffc}
\end{figure}

Results for the difference profiles, 
$\delta x_\ell(t) \equiv x_\ell(t)-x_\ell^\ast$
and the similarly defined $\delta y_\ell(t)$, for the nanotube
with $p=q=1$, $\alpha=2(\sqrt{2}-1)$, $\beta=2-\sqrt{2}$
are exhibited in Fig.~\ref{fig:pq1diffc}. Late-time data were taken at 
$t=500$ (for comparison, the corresponding steady-state densities shown 
in Fig.~\ref{fig:pq1altc} were taken at $t=2500$).

It is seen that the spatial dependence of  $\delta x_\ell(t)$ and
$\delta y_\ell(t)$ is indeed very close to that anticipated in
Eqs.~(\ref{eq:xellnu}),~(\ref{eq:yellnu}), although the numerical
results show a slight skew. 
The fit to a sine form shown as a dashed line in Fig.~\ref{fig:pq1diffc}  
corresponds to $\chi^2_{\rm dof}=49$, which is unsatisfactory.
We then allowed for a nonzero gap, by returning to the more general 
expressions Eqs.~(\ref{eq:xt}) and~(\ref{eq:yt}).
Fitting to the $n=1$ term of Eq.~(\ref{eq:xt}), i.e.,
\begin{equation}
\delta x_\ell(t)= -a(t)\,e^{\phi\ell}\,\sin\left(\frac{\pi\ell}{L}\right)\ ,
\label{eq:abfit}
\end{equation}
we found the full-line curve depicted in Fig.~\ref{fig:pq1diffc}, with $\phi=-0.022(1)$,
$\chi^2_{\rm dof}=3.2$. The small, but definitely non-zero, estimate of $\phi$ is in line
with the steady-state results shown in Fig.~\ref{fig:pq1altc} in that
both indicate the approximate, rather than exact, character of the mean-field description
for $p \neq 2q$. 

Furthermore, the difference profiles are almost entirely 
sublattice-independent, a feature which is not obviously
forthcoming from the theory of Sec.~\ref{sec:pneq2q}.
It can be shown (see Appendix~\ref{sec:app1}) that this results from
the existence of two distinct relaxation rates: one which is 
very fast, size-independent [$\,$which brings the sublattice profiles
to shapes rather close to their steady-state ones$\,$] and a slower one,
with characteristic times of the usual $L^z$ form. In the 
(not very short)-time regime for which the latter applies
the sublattice distinction disappears for difference profiles, and 
the dynamics can be described in an effective continuum approximation 
through linear equations resulting from a Cole-Hopf transformation.
For example,
difference profiles taken at $t=250$
for the system considered in Fig.~\ref{fig:pq1diffc} 
already exhibit a degree of sublattice-independence very similar 
to that shown in the Figure. Using 
Eqs.~(\ref{eq:xellnu}),~(\ref{eq:yellnu}) for simplicity, 
defining $G^{\prime\prime}(t) \equiv G^\prime\,e^{-\lambda_-(\pi/L)t}$ 
one finds by fitting numerical data $G^{\prime\prime}(250)/
G^{\prime\prime}(500)  \approx 4.4$, which corresponds
to $\lambda_-(\pi/L) \approx 6 \times 10^{-3}$. Direct evaluation
via the theoretical prediction Eq.~(\ref{eq:lambdamin}), using the
mean field values for $x^\ast$, $y^\ast$, $r$ from 
Eqs.~(\ref{eq:xbarybar}) and~(\ref{eq:rsdef})
gives $\lambda_-(\pi/L)=1.0\times 10^{-3}$.

Turning now to non-critical systems, 
proceeding along the lines followed
above one can again adapt  Eqs.~(\ref{eq:xt}),~(\ref{eq:yt}) to make
allowance for the position dependence of steady state profiles,
for systems away from criticality but with $\alpha$ and 
$\beta$ obeying Eq.~(\ref{eq:newcond}). 

\begin{figure}
{\centering \resizebox*{3.3in}{!}{\includegraphics*{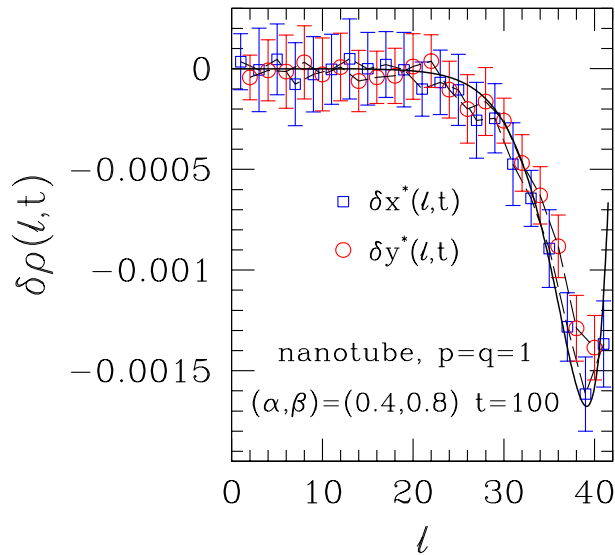}}}
\caption{
Late-time difference profiles, $\delta x_\ell(t) 
\equiv x_\ell(t)-x_\ell^\ast$, and similarly for $\delta y_\ell(t)$,
against position along flow direction
for nanotube $p=q=1$ at $\alpha=0.4$, $\beta=0.8$,
for $t=100$. The full line is the fit of  $\delta x_\ell(t)$ to a sine 
plus exponential form, using only the $n=1$ term of
Eq.~(\ref{eq:xt}) [$\,$in an adapted form to allow for the position
dependence of steady state profiles, see text$\,$].
}
\label{fig:pq1da4b8}
\end{figure}

For $\alpha=0.4$, $\beta=0.8$ the numerically-obtained steady state
profiles turned out to be nearly flat down to $3-4$ parts in $1000$, 
 with $x_\ell \approx 0.324$, $y_\ell \approx 0.209$, except very near 
the system's ends. These values are 
rather close to the mean field ones  predicted via
Eq.~(\ref{eq:xbarybar}), namely $x_\ell=1/3$, $y_\ell=1/5$~.
The late-time difference profiles obtained in the way described above,
at $t=100$, are displayed in Fig.~\ref{fig:pq1da4b8}.
Fitting to Eq.~(\ref{eq:abfit})
gives a fairly good account of the behavior of $\delta x_\ell(t)$
against $\ell$; also, the sublattice independence of difference
profiles is obeyed to a good extent, though some slight discrepancies
remain near the ejection end. From 
Eqs.~(\ref{eq:xbarybar}),~(\ref{eq:xt})--(\ref{eq:Siqn}), theory 
predicts that the coefficient $\phi$ in the position dependence of 
late-time density profiles should be $\phi=(\ln 8)/2=
1.04 \dots$, and that for the time dependence the slowest decay 
rate should be $\lambda_1^T=0.166 \dots$. 

The fitting curve shown in  
Fig.~\ref{fig:pq1da4b8} corresponds to $\phi=0.34(1)$. A measure of 
self-consistency of the latter can be gained by pointing out that, 
if $\phi L \gtrsim 5-6$ the minimum of Eq.~(\ref{eq:abfit}) is located at
$\ell \approx L-(1/\phi)$. Visual inspection of Fig.~\ref{fig:pq1da4b8}
confirms that numerical data indeed behave in this way. On the other
hand, the mismatch between predicted and observed values of $\phi$ 
is a rather extreme illustration of the limitations of mean field
mapping predictions for $p \neq 2q$, already evident e.g. in
the density profiles of Fig.~\ref{fig:pq1altc}.

We checked the theoretical prediction for $\lambda_1$ by comparing
difference profiles at $t=80$ with those for $t=100$. Referring to 
Eq.~(\ref{eq:abfit}), one gets $a(100)/a(80)=0.08 \pm 0.05$, 
broadly compatible with 
$e^{-20\,\lambda_1^T}=0.03615 \dots$. 

\subsection{Factorization in steady state} 
\label{sec:fac}

It was seen in Sec.~\ref{sec:num:pneq2q} that numerical results
for steady state density profiles on nanotubes and alternating-bond
chains with $p \neq 2q$ are at variance with the predictions of
mean field Mobius mapping.
Mismatches of similar order have been found between mean-field results
and numerical work regarding steady-state currents in graphene-like 
structures with $p \neq 2q$~\cite{hex13}.

In the following, we expand on comments made in Ref.~\onlinecite{hex13},
regarding the issue of factorization in steady state.

It is known for the strictly one-dimensional TASEP that, along
$\alpha+\beta=1$ the correlations vanish, i.e., the probabilities for
occupation variables on different sites factorize~\cite{derr93}.
As a consequence of this, along that line the mean field mapping 
produces exact results. For nanotubes one 
can then check for factorization (or its absence), 
in order to test the extent 
to which the predictions given via Mobius mappings are accurate. 

A direct test can be implemented by considering the (connected)
correlation function, 
\begin{equation}
C_{ij} \equiv \langle J_{ij}\rangle - p_{ij}\, \langle \tau_i 
\rangle\,\left( 1 - \langle\tau_j\rangle\right)\ ,
\label{eq:c_ij}
\end{equation}
where $\langle J_{ij}\rangle$ is the average current across a chosen bond 
$ij$  with rate $p_{ij}$, 
connecting sites $i$, $j$ with respective mean occupations 
$\langle \tau_i \rangle$, $\langle\tau_j\rangle$. 
Factorization then corresponds to $C_{ij} \equiv 0$ for all bonds $ij$.

\begin{figure}
{\centering \resizebox*{3.3in}{!}{\includegraphics*{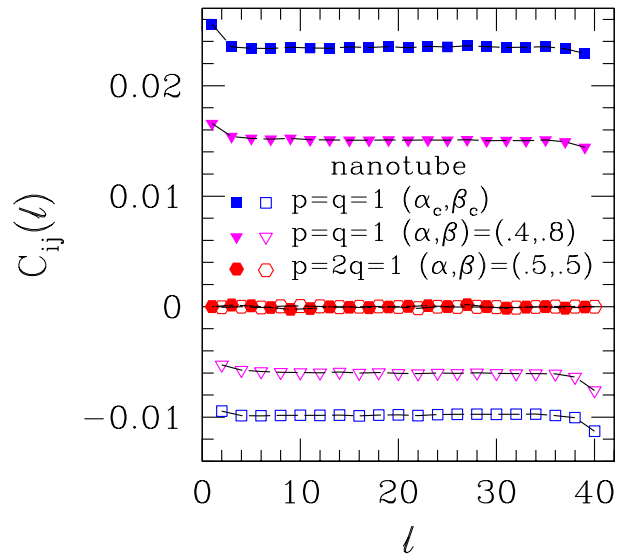}}}
\caption{
Nanotube with $N_w=14$, $N_r=10$: $C_{ij}$ of eq.~(\ref{eq:c_ij}),
averaged over transverse coordinate, against position along flow 
direction. Full symbols: $x-$ sublattice. Empty symbols: $y-$ sublattice.
For  $p=q=1$, $(\alpha_c,\beta_c)=(2(\sqrt{2}-1),2-\sqrt{2})$.
}
\label{fig:fact1}
\end{figure}

We have found that for the nanotube with $p=1$, $q=1/2$ $C_{ij}$ vanishes 
to the accuracy of  simulation (typically $1$ part  in $10^5$) 
on (and only on) the line $\alpha+\beta=1$, the same as in the strictly
one-dimensional case. This is a non-trivial higher-dimensional
generalization of a well known result for the linear chain. On the other
hand, with  $p=1=q=1$, we followed the predicted factorization
line, Eq.~(\ref{eq:newcond}), and found that in simulations of similar
accuracy, the factorization is no better than $1$ part in $10^2$. 
This is illustrated in Fig.~\ref{fig:fact1}, where data taken at the 
respective  predicted critical points, namely $\alpha=\beta=1/2$ 
[$\,p=2q=1\,$] and
$\alpha=2(\sqrt{2}-1)$, $\beta=2-\sqrt{2}$ [$\,p=q=1\,$] are shown.
For $p=q=1$, data are shown also for $(\alpha,\beta)=(0.4,0.8)$, i.e.,
further along the predicted factorization line Eq.~(\ref{eq:newcond}).

Still with  $p=q=1$ we thoroughly scanned the $(\alpha,\beta)$
parameter space, and found no evidence either of uniform sublattice 
profiles or of vanishing of $C_{ij}$.

\section{Discussion and Conclusions} 
\label{sec:conc}

We have presented a mean-field theory for the dynamics of driven flow 
with exclusion in graphene-like structures, and numerically checked 
its predictions.

For the special combination of bond rates $p=2q$ in the
nanotube geometry, Eqs.~(\ref{eq:jmf1})--(\ref{eq:nstrho2})
show that the sublattice distinction goes away in mean field. 
So a continuum picture can apply, giving Eq.~(\ref{eq:contdyn}) 
for which a time-dependent solution is found by using the Cole-Hopf 
transformation.

For the special boundary rates $\alpha=\beta=1/2$ which corresponds to criticality
in the 1D chain with uniform rates, predictions for the early-- and late-time
behavior of density profiles are made respectively in Eqs.~(\ref{eq:early})
and~(\ref{eq:late}). These are borne out by numerics to very good
accuracy, see Figs.~\ref{fig:in_slope} and~\ref{fig:sinefit}.
We focused on late-time behavior, for both 1D and nanotube geometries,
and showed that by systematically analyzing the results of density
profile fits  to Eq.~(\ref{eq:late}) it was possible
[$\,$see Eqs.~(\ref{eq:late2})--~(\ref{eq:cprime})$\,$] to extract rather
accurate estimates of the dynamic exponent $z$. For strictly 1D systems,
we find $z=1.51(1)$, in excellent agreement with the anomalous value
$z=3/2$ which is known~\cite{derr98,sch00,rbs01} to apply in that case. 
For nanotubes, we found strong indications (see Fig.~\ref{fig:ntfit})
that the limiting behavior for very long length
depends on whether one considers (quasi-1D) systems of fixed width, or
square-like ones with constant aspect ratio; while the former
exhibit $z$ again close to $3/2$, the latter are characterized by
$z$ consistent with the mean-field value of $2$ (within error bars).  
In the standard language of critical phenomena, this would
mean that the upper critical dimensionality for TASEP dynamics is
certainly $D_c \leq 2$.
 
On the factorization line $\alpha+\beta=1$ where
steady-state profiles are uniform both for uniform-rate chains and nanotubes
with $p=2q$~\cite{hex13}, we took $\alpha=0.3$, away from criticality. 
The main distinguishing feature here, relative to the critical case, 
is the opening of a gap of amplitude $\lambda^2/2=(1-2\alpha)^2/2$, 
associated with the characteristic  length $\lambda^{-1}$. The 
predicted effects of this on late-time profile shapes are spelt out 
in Eq.~(\ref{eq:rhoab1b}), which is qualitatively supported by numerical data
(see Fig.~\ref{fig:expsinefit}). 

However, the quantitative effects, on the density profiles,
of having $\lambda \neq 0$ are not accurately described by the present mean-field theory.
Partly because of this, attempts to extract the dynamical exponent $z$, by
procedures similar to those followed in the gapless case, met
with the difficulties described in Sec.~\ref{sec:num_peq2qa3b7}.

We then resorted to an overall consistency check, based on keeping fixed
the effective exponent value $z^\prime=1$ which holds for the low-current
phase in 1D systems~\cite{dqrbs08}. The resulting fits of numerical estimates 
of the coefficients  appearing in the exponential time decay factor of 
Eq.~(\ref{eq:rhoab1b}), shown in Fig.~\ref{fig:a3b7ntfit}, produce a
reasonably self-consistent picture.

For nanotubes with $p \neq 2q$ (and chains with alternating bonds), 
Fig.~\ref{fig:pq1altc} illustrates that  predictions for steady state 
profiles from mean field mapping are not as accurate as for $p=2q$, or
for uniform chains. In particular, numerically-generated profiles 
display a distinctive degree of nonuniformity along the predicted 
factorization line.

Since dynamics concerns the evolution from initial to steady state, 
rather than
the detailed (time-independent) properties of the latter, we adapted 
our original formulation to allow for the observed non-uniformity
of the sublattice-dependent limiting profile shapes, see 
Eqs.~(\ref{eq:xellnu}),~(\ref{eq:yellnu}) for critical systems,
and Eq.~(\ref{eq:abfit}) for the off-critical case. We found that
for late times the difference profiles thus defined behave in a 
very close 
way to that predicted by the theory of Sec.~\ref{sec:pneq2q}, see 
respectively Figs.~\ref{fig:pq1diffc} and~\ref{fig:pq1da4b8}.

This latter remark deserves to be qualified, inasmuch as it refers
strictly to the functional forms displayed in 
Eqs.~(\ref{eq:xellnu}),~(\ref{eq:yellnu}), or Eq.(~\ref{eq:abfit})
[$\,$respectively sine, or sine plus exponential$\,$] rather than
to numerical values of the associated parameters [$\,$respectively
$G^\prime$, $H^\prime$, or $a(t)$, $\phi\,$] which we estimate via
best-fitting procedures. Although this is not as stringent a test
of mean field theory as would be the case if the theory-predicted
parameter values were used, working this way allows one to separate 
possible shortcomings of the  mean field approximation in functional forms
versus those in parameters. Furthermore, one can have a quantitative estimate 
(through $\chi^2$ values) of discrepancies in mean field functional forms, 
rather than the qualitative impressions from the comparisons with 
full predictions coming from theory; one can also get quantitative estimates of 
parameters affected by fluctuation effects absent from mean field theory, 
with the hope that modest generalizations (like domain wall theory) 
might more accurately provide such parameters. 

An additional feature of the late-time difference profiles 
is that they are almost entirely
sublattice-independent. This property 
has been shown (see Appendix~\ref{sec:app1}) to result from               
the coexistence of two distinct relaxation rates: a very fast,
size-independent one, and a slower one 
with characteristic times of the usual $L^z$ form. 
The latter applies, within an effective continuum picture, to the Goldstone modes
resulting from particle number conservation.
If one accepts that an accurate description of TASEP via mean field mapping
goes together with full applicability of a continuum approximation, 
this would then explain why the late-time density differences
generally fall in line with mean-field, continuum-like, predictions.

Detailed comparison of theoretical predictions from Sec.~\ref{sec:pneq2q}
to numerical results beyond overall profile shapes turns out to not be
as accurate as for $p=2q$. For the system considered in 
Fig.~\ref{fig:pq1diffc} theory gives for the exponential 
time-decay coefficient of Eq.~(\ref{eq:lambdamin})
$\lambda_-(\pi/L)=1 \times 10^{-3}$, while adjusting to numerical data  
gives $\lambda_-(\pi/L) \approx 6  \times 10^{-3}$. Similarly, 
for the non-critical system corresponding
to Fig.~\ref{fig:pq1da4b8}, using the theoretical prediction for
$\lambda_1$ of Eqs.~(\ref{eq:xt}),~(\ref{eq:yt}) would give a
ratio of difference-profile coefficients at $t=100$ and $t=80$ equal to
$0.03615 \dots$, while this same ratio is estimated from numerical data as
$0.08 \pm 0.05$.  

Finally, in Sec.~\ref{sec:fac} we showed that a direct test of
factorization of correlation functions in steady state produces a
clear correspondence between uniformity of observed
steady state profiles, on the one hand, and numerical evidence of
vanishing of correlations, on the other.

\begin{acknowledgments}
We thank Fabian Essler for helpful discussions.
S.L.A.d.Q. thanks the Rudolf Peierls Centre for Theoretical Physics, Oxford, for
hospitality during his visit. The research of S.L.A.d.Q. is supported by 
the Brazilian agencies CNPq  (Grant No. 303891/2013-0), 
and FAPERJ (Grants Nos. E-26/102.760/2012 and E-26/110.734/2012).
\end{acknowledgments}
 
\appendix

\section{Fast transient equalization of sublattices}
\label{sec:app1}
For our purposes here, it is convenient to adopt
the following notation: sites on the $x-$ sublattice have even
site label, with mean occupation $x_{2\ell}$; for the $y-$
sublattice, with odd labels one has the mean occupation $y_{2\ell+1}$.
Bond rates are $p$ and $p^\prime \equiv 2q$.

Thus the mean field defining equations for
currents and occupations, and their evolution,
Eqs.~(\ref{eq:jmf1})--(\ref{eq:nstrho2}), become
\begin{eqnarray}
J_{2\ell\,2\ell+1}=p\,x_{2\ell}\,(1-y_{2\ell+1})\ \ \,
\label{eq:jmf1b}\\
K_{2\ell-1\,2\ell}=p^\prime\,y_{2\ell-1}\,(1-x_{2\ell})\ .
\label{eq:jmf2b}
\end{eqnarray}
\begin{eqnarray}
{\dot x}_{2\ell} = K_{2\ell-1\,2\ell} - J_{2\ell\,2\ell+1}\ \ \,
\label{eq:nstrho1b}\\
{\dot y}_{2\ell+1} = J_{2\ell\,2\ell+1} - K_{2\ell+1\,2\ell+2}\ .
\label{eq:nstrho2b}
\end{eqnarray}
Eqs.~(\ref{eq:nstrho1b}) and~(\ref{eq:nstrho2b}) give:
\begin{equation}
\frac{\partial}{\partial t}\,\left(x_{2\ell}+ y_{2\ell+1}\right) =
 K_{2\ell-1\,2\ell} - K_{2\ell+1\,2\ell+2}\ .
\label{eq:xplusy}
\end{equation}
When a continuum picture applies, the right-hand side of
Eq.~(\ref{eq:xplusy}) becomes like a space derivative of $K$ and is
then small, so $x_{2\ell}+ y_{2\ell+1}$ becomes a slow variable;
similarly for $y_{2\ell-1}+x_{2\ell}$.

On the other hand, any linear combination $a\,x_{2\ell}+ b\,y_{2\ell+1}$
with $a \neq b$ decays rapidly towards zero. This implies that the
rapid decay is towards "adiabatic" values of $x_{2\ell}$, $y_{2\ell+1}$
such that all $K_{2\ell-1\,2\ell} - J_{2\ell\,2\ell+1}$ and
$J_{2\ell\,2\ell+1} - K_{2\ell+1\,2\ell+2}$ are zero. That is,
\begin{equation}
K_{2\ell-1\,2\ell}=J_{2\ell\,2\ell+1}=J_{2\ell\,2\ell+1}=K_{2\ell+1\,2\ell+2}
= \cdots = C(t)\ .
\label{eq:adiab}
\end{equation}
The function $C(t)$ is the adiabatically evolving "conserved current"
related to the particle conservation represented by the set of
equations Eq.~(\ref{eq:xplusy}) for all $\ell$.
Those equations determine the adiabatic evolution of the conserved
densities.

After the very fast transients have died out the profiles on the two
sublattices still differ from their steady-state values $\bar{x}_{2\ell}$,
$\bar{y}_{2\ell+1}$ by amounts
$\delta x_{2\ell}(t)$, $\delta y_{2\ell+1}(t) $; as shown in the
following, such differences are essentially the same for either
sublattice, as their approach
to zero is governed by a single continuum-like evolution equation.

The fast time scales for the evolution of $a\,x_{2\ell}+ b\,y_{2\ell+1}$
with $a \neq b$, coming from equations without nearly cancelling currents,
and so without conserved or spatial derivative aspects, have rates set
just by $p$ and $p^\prime$, and not by wave vectors or system size $L$.
So they are of order one, rather than a power of $L$ or wavelength.

In the subsequent evolution (after the initial transient regime)
(i)\ we can interpolate the density variables between
the sites of their sublattice, making very little error; and
(ii)\ use the resulting "continuumization"
of sites to find the conserved current differences in terms of
spatial derivatives: e.g., $\widetilde{y}_{2\ell}$ is the
interpolation of the odd sublattice variables $y_{2\ell-1}$,
$y_{2\ell+1}$; similarly for $\widetilde{x}_{2\ell+1}$. So,
\begin{eqnarray}
K_{2\ell-1\,2\ell}-K_{2\ell+1\,2\ell+2}=\qquad\qquad
\nonumber\\
=p^\prime\,\left[\,y_{2\ell-1}
(1-x_{2\ell})-y_{2\ell+1}(1-x_{2\ell+2})\right] \approx \qquad\qquad
\nonumber\\
 \approx p^\prime\,\left(-2\frac{\partial}{\partial \ell}\right)\,\left[\,
\widetilde{y}_{2\ell}(1-\widetilde{x}_{2\ell+1})\right]\ ,\qquad\qquad
\label{eq:interp}
\end{eqnarray}
and similarly for differences of adjacent $J^\prime$s.

Combining Eqs.~(\ref{eq:xplusy})  and~(\ref{eq:interp})
[$\,$and their counterparts for
$y_{2\ell-1}+x_{2\ell}$ and $J_{2\ell-2\,2\ell-1}-J_{2\ell\,2\ell+1}$,
respectively$\,$], omitting the subscripts and tilde signs, redefining
$\ell$ as an "average" coordinate shared by a pair of adjacent $x-$ and
$y-$ subllattice sites, and defining $\rho(\ell)=\frac{1}{2}(x_\ell+
y_\ell)$, one gets:
\begin{equation}
\frac{\partial \rho}{\partial t}=-\left(\frac{p+p^\prime}{2}\right)\,
\frac{\partial}{\partial\ell}\left[\rho(1-\rho)-\frac{1}{2}
\frac{\partial\rho}{\partial\ell}\right]\ .
\label{eq:c-h2}
\end{equation}
This is now the form which the Cole-Hopf transformation linearizes.


\begin{thebibliography}{99}
\bibitem{hex13} R. B. Stinchcombe, S. L. A. de Queiroz, M. A. G. Cunha, and
Belita Koiller, \pre {\bf 88}, 042133 (2013). 
\bibitem{derr98} B. Derrida, Phys. Rep. {\bf 301}, 65 (1998).
\bibitem{sch00} G. M. Sch\"utz, in {\it Phase Transitions and
Critical Phenomena}, edited by C. Domb and J. L. Lebowitz (Academic,
New York, 2000), Vol. 19.
\bibitem{mukamel} B. Derrida, E. Domany, and D. Mukamel, J. Stat. Phys. {\bf 69},
667 (1992). 
\bibitem{derr93} B. Derrida, M. Evans, V. Hakim, and V. Pasquier, 
J. Phys. A {\bf 26}, 1493 (1993).
\bibitem{rbs01} R. B. Stinchcombe, Adv. Phys. {\bf 50}, 431 (2001).
\bibitem{be07} R. A. Blythe and M. R. Evans, J. Phys. A {\bf 40}, R333 (2007).
\bibitem{cmz11} T. Chou, K. Mallick, and R. K. P. Zia, Rep. Prog. Phys. 
{\bf 74}, 116601 (2011).
\bibitem{sz95} B. Schmittmann and R. K. P. Zia, in {\it Phase Transitions and
Critical Phenomena}, edited by C. Domb and J. L. Lebowitz (Academic,
New York, 1995), Vol. 17.
\bibitem{rb02} R. Bundschuh, \pre {\bf 65}, 031911 (2002).       
\bibitem{kvo10} T. Karzig and F. von Oppen, \prb {\bf 81}, 045317 (2010).
\bibitem{rsss98}  N. Rajewsky, L. Santen, A. Schadschneider, and
M. Schreckenberg, J. Stat. Phys. {\bf 92}, 151 (1998).
\bibitem{dqrbs08} S. L. A. de Queiroz and  R. B. Stinchcombe, 
\pre {\bf 78}, 031106 (2008).
\bibitem{RMP} J-C. Charlier, X. Blase, and  S. Roche, 
\rmp {\bf 79}, 677 (2007); A. H. Castro Neto, F. Guinea, N. M. R. Peres, 
K. S. Novoselov, and A. K. Geim, {\it ibid.} {\bf 81}, 109 (2009).
\bibitem{hopf50} E. Hopf, Commun. Pure Appl. Math. {\bf 3}, 201 (1950).
\bibitem{cole51} J. D. Cole, Q. Appl. Math. {\bf 9}, 225 (1951).
\bibitem{bn}  M. P. Nightingale and  H. W. J. Bl\"ote,  J. Phys. A {\bf 15},
L33 (1982); H. W. J. Bl\"ote and M. P. Nightingale,  Physica A {\bf 112},
405 (1982); {\it ibid}, {\bf 134}, 274 (1985).
\bibitem{dq12} S. L. A. de Queiroz, \pre {\bf 86}, 041127 (2012).
\bibitem{unpub}  R. B. Stinchcombe and  S. L. A. de Queiroz, unpublished.
\bibitem{ksks98} A. B. Kolomeisky,  G. M. Sch\"utz, E. B. Kolomeisky,
and J. P. Straley, J. Phys. A {\bf 31}, 6911 (1998).
\bibitem{ps99} V. Popkov and G. M. Sch\"utz, Europhys. Lett. {\bf 48}, 257 (1999).
\bibitem{ds00} M. Dudzinsky and G. M. Sch\"utz, J. Phys. A {\bf 33}, 8351 (2000).
\bibitem{ess05} J. de Gier and F. H. L. Essler, \prl {\bf 95}, 240601
(2005); J. Stat. Mech.: Theory Exp. (2006) P12011.

\end{thebibliography}
\end{document}